\documentclass[trackchanges,twocolumn]{aastex701}

\usepackage{amsmath}
\usepackage{tabularx}
\usepackage{makecell}
\usepackage{adjustbox}
\usepackage{subcaption}
\definecolor{orcidlogocol}{HTML}{A6CE39}
\newcommand{\orcidicon}[1]{\href{https://orcid.org/#1}{\textcolor{orcidlogocol}{\aiOrcid}}}

\begin{document}

\title{Magnetic Configuration Imprints on Quasi-Periodic Variability in GRMHD Simulations of Thin Accretion Disks}

\author[0009-0004-8669-2411]{Jing-Ze Xia} 
\affiliation{Tsung-Dao Lee Institute, Shanghai Jiao Tong University,
1 Lisuo Road, Shanghai 201210, PR China}
\email{jackxia@sjtu.edu.cn}

\author[0000-0003-0292-2773]{Hong-Xuan Jiang}
\affiliation{Tsung-Dao Lee Institute, Shanghai Jiao Tong University,
1 Lisuo Road, Shanghai 201210, PR China}
\email{hongxuan_jiang@sjtu.edu.cn}

\author[0000-0002-4064-0446]{Indu K. Dihingia}
\affiliation{Institute of Fundamental Physics and Quantum Technology, School of Physical Science and Technology, Ningbo University, Ningbo, Zhejiang 315211, People’s
Republic of China}
\affiliation{Tsung-Dao Lee Institute, Shanghai Jiao Tong University,
1 Lisuo Road, Shanghai 201210, PR China}
\email{ikd4638@gmail.com, ikd4638@sjtu.edu.cn}

\author[0000-0002-8131-6730]{Yosuke Mizuno}
\affiliation{Tsung-Dao Lee Institute, Shanghai Jiao Tong University,
1 Lisuo Road, Shanghai 201210, PR China}
\affiliation{School of Physics and Astronomy, Shanghai Jiao Tong University,
800 Dongchuan Road, Shanghai 200240, PR China}
\affiliation{Key Laboratory for Particle Physics, Astrophysics and Cosmology (MOE),
Shanghai Key Laboratory for Particle Physics and Cosmology,
Shanghai Jiao Tong University, 800 Dongchuan Road, Shanghai 200240, PR China}
\affiliation{Institut f\"ur Theoretische Physik, Goethe-Universit\"at Frankfurt,
Max-von-Laue-Str.~1, D-60438 Frankfurt am Main, Germany}
\email{mizuno@sjtu.edu.cn}

\begin{abstract}
The origin of quasi-periodic oscillations (QPOs) in black hole accretion flow remains uncertain, particularly regarding the role of magnetic field configurations in shaping disk structure and variability signatures. We investigate this using global two- and three-dimensional (2D and 3D) general relativistic magnetohydrodynamic (GRMHD) simulations of geometrically thin disks initialized with different multi-loop magnetic field configurations. These configurations naturally produce a puffed-up inner region. We find that QPO-like variability arises in the effective viscosity and mass accretion rate, with frequencies following the local radial epicyclic frequency and its harmonics. Time-series diagrams show coherent, inclined stripe-like patterns associated with inertial–acoustic perturbations, while power spectra exhibit narrow bands of enhanced variability linked to truncation radii associated with magnetic fields. Cross-correlation analysis reveals a finite lag between pressure and Maxwell stress at these interfaces, consistent with viscous–epicyclic overstability. The magnetic topology regulates both the truncation radius and the location of resonant cavities that sustain oscillations. As the disk becomes thicker, increased turbulent diffusion suppresses the overstability and the associated QPO signals. We find that the QPO frequency ranges and their evolution are consistent with observations of black hole X-ray binaries during outbursts. These results suggest that magnetic field configurations play a pivotal role in shaping disk structure and variability in accreting black holes.
\end{abstract}



\section{Introduction}\label{sec:intro}

Accretion onto compact objects is among the most efficient channels for converting gravitational potential energy into radiation in the universe. The classical thin-disk model \citep{1973A&A....24..337S} and its general relativistic extension \citep{1973blho.conf..343N} successfully explain the thermal spectra of black-hole X-ray binaries (BHXRBs) in their high/soft states and many features of luminous active galactic nuclei \cite[AGNs; e.g.,][]{2006ARA&A..44...49R,2007A&ARv..15....1D}.  In these models, angular momentum transport is parameterized by a dimensionless viscosity coefficient $\alpha$, which encapsulates complex microphysics into an effective stress proportional to the local pressure. Although this phenomenological framework effectively captures global energetics and spectra, the physical origin and underlying structure of the stress remain poorly understood. Furthermore, it cannot account for the rapid variability observed in many accreting systems, particularly the quasi-periodic oscillations (QPOs).

The physical mechanism underlying angular-momentum transport is now understood as triggered by magnetorotational instability \cite[MRI;][]{1991ApJ...376..214B,1998RvMP...70....1B}, which taps the free energy of differential rotation to drive MHD turbulence and an effective Maxwell stress. Global general-relativistic MHD (GRMHD) simulations have therefore become a standard tool for studying black-hole accretion flow across a wide range of geometries, from geometrically thick, radiatively inefficient flows to thinner, radiatively efficient disks \cite[e.g.,][]{2003ApJ...589..444G,2012MNRAS.423.3083M,2012MNRAS.426.3241N,2011ApJS..197...31S,2019MNRAS.487..550L,2022Univ....8...85M,Dihingia2025}.  Most numerical work, however, has focused either on hot, geometrically thick accretion flows or on magnetically arrested disks \cite[MADs;][]{2011MNRAS.418L..79T,2012MNRAS.423.3083M}, in which a large-scale vertical field accumulates near the black hole and strongly regulates the dynamics.  In contrast, Standard And Normal Evolution (SANE) disks, in which the net vertical flux remains modest \citep{2020MNRAS.495.1549N}, have received comparatively less attention in the geometrically thin regime, despite their closer resemblance to the classical Novikov-Thorne picture.

From an observational perspective, QPOs in BHXRBs provide a powerful probe of strong-gravity accretion flow \cite[see][for reviews]{2006csxs.book...39V,2006ARA&A..44...49R,2014SSRv..183...43B}.  High-frequency QPOs (HFQPOs) display relatively stable frequencies that often scale with the orbital or epicyclic frequencies in the inner disk, suggesting a dynamical origin tied to diskoseismic modes, relativistic precession, or other oscillatory phenomena \cite[e.g.,][]{1999PhR...311..259W,2001PASJ...53....1K,1991ApJ...378..656N,1997ApJ...476..589P,1998ApJ...492L..59S,2009MNRAS.397L.101I,2008MNRAS.387..446T,2025ApJ...982L..21D,2025ApJ...995..112J}. Among the proposed mechanisms, viscous-epicyclic overstability has emerged as a particularly attractive candidate for exciting inertial-acoustic (p-mode) oscillations in thin disks \citep{1978MNRAS.185..629K,2006MNRAS.372.1829L,2009MNRAS.393..979L,2015MNRAS.446..240M}. In this picture, the key ingredient is a finite phase lag between pressure perturbations and the viscous (or turbulent) stress; if the stress responds with an appropriate delay, it can extract energy from differential rotation and do net positive work on epicyclic oscillations, rendering them overstable rather than damped.

Linear analyses have shown that viscous overstability can operate in thin, nearly Keplerian disks over a range of radii and wavenumbers \citep{2006MNRAS.372.1829L,2015MNRAS.446..240M,2022MNRAS.510..146H}. Local MRI simulations have demonstrated that turbulent stresses can indeed exhibit systematic lag related with pressure fluctuations \citep{2022MNRAS.510..146H}.  However, it remains unclear to what extent this mechanism survives in fully nonlinear, global GRMHD flows with realistic magnetic topology. In particular, most existing studies of viscous overstability either adopt prescribed $\alpha$ viscous laws or focus on local shearing boxes, leaving open the question of how global field geometry, disk thickness, and radial inhomogeneities (such as density or magnetization interfaces) influence the excitation, trapping, and saturation of overstable modes in geometrically thin SANE disks.

Magnetic topology plays a dual role in this problem. On the one hand, it sets the amplitude and spatial distribution of the turbulent stress that underlies the effective $\alpha$ parameter, thereby influencing the viscous driving and damping. On the other hand, multi-loop magnetic configurations naturally create a radial structure in density, magnetization, and effective viscosity, including transition layers and cavities that may act as resonant trapping regions for inertial-acoustic waves \citep{2020MNRAS.495.1549N,2021MNRAS.505.3596D,2022MNRAS.517.5032D}.   It is therefore timely to ask whether realistic SANE-like magnetic geometries can both modify the inner disk structure (e.g., by producing local puff-up and truncation) and at the same time seed or sustain QPO-like oscillations via viscous-epicyclic overstability.

In this work, we address these questions using high-resolution GRMHD simulations of geometrically thin, SANE accretion disks around a spinning black hole. We initialize the disk with a Novikov-Thorne-like density profile and thread it with a set of magnetic loops of varying wavelength, following the multi-loop prescription of \citet{2020MNRAS.495.1549N}. By systematically varying the loop size, we control the degree of inner-disk puff-up and the location of density/magnetization interfaces.
Beyond characterizing the global disk structure and jet power, we focus on the temporal variability of the flow, identifying QPO-like features associated with both global inertial-acoustic modes near the radial epicyclic frequency and localized oscillations anchored at loop interfaces. To directly diagnose the mechanism driving these oscillations, we measure at each radius the cross-correlation between pressure and Maxwell stress and use it as a time-domain proxy for the stress-pressure phase lag, thereby connecting global GRMHD turbulence to the viscous-epicyclic overstability framework developed in analytic and local studies.

The paper is organized as follows. In Section~\ref{sec:setup}, we describe our numerical methods, initial conditions, and magnetic field configurations. Section~\ref{sec:result} presents the global evolution of the simulated disks. In Section~\ref{time-var} and Section~\ref{3D}, we analyze the QPO-like signals and their connection to inertial-acoustic modes and viscous overstability.  We summarize our main results and discuss their implications in Section~\ref{sec:summary}.

\section{initial setup} \label{sec:setup}

All simulations in this work are performed with the GPU-accelerated GRMHD code \texttt{KHARMA}, a GPU-enabled extension of \texttt{iharm3D} \citep{2024arXiv240801361P}. Both \texttt{KHARMA} and \texttt{iharm3D} descend from the original \texttt{HARM} scheme, which solves the equations of ideal magnetohydrodynamics in a general relativistic framework \citep{2003ApJ...589..444G}.

%

In our simulations we adopt a \textit{funky modified Kerr-Schild (FMKS)} coordinate system to concentrate grid resolution toward the equatorial plane and to relax the Courant time-step constraint near the polar axis \citep{2025arXiv250717818C}. 

We initialize a geometrically thin accretion disk following \citet{2021MNRAS.505.3596D}; the details of this model are summarized in the Appendix~\ref{appenA}. The disk is threaded by a multi-loop, SANE-like magnetic field configuration inspired by \citet{2020MNRAS.495.1549N}, designed to mimic a more dynamo-like field (see Appendix~\ref{appenB} for details). The parameters we used are summarized in Table.~\ref{Model comparison}.
\begin{table*}
\caption{Model parameters including $\lambda_r$, radial component, resolution, and simulation time.}
\label{tab:model_parameters}
\centering

\begin{tabular}{cccccc}
\hline\hline
Model & $\lambda_r$ & Radial component $r^w$ & Resolution & $t_\text{end}\,(\rm M)$ & note\\
\hline
A & 20 & $w=0$ & $1024\times768$ & 38,000 & small loop\\
B & 40 & $w=0$ & $1024\times768$ & 38,000 & midium loop\\
C & 60 & $w=0$ & $1024\times768$ & 38,000 & large loop\\
D & 20 & $w=0.5$ & $1024\times768$ & 25,000 & small loop with $r^{0.5}$\\
E & 20 & $w=1$ & $1024\times768$ & 25,000 & small loop with $r^{1}$\\
\tt{Th3D20} & 20 & $w=0$ & $384\times192\times192$ & 18,000 & 3D case of model~A\\
\tt{Th3D05} & 5 & $w=0$ & $384\times192\times192$ & 18,000& smaller loop\\
\hline
\end{tabular}

\end{table*}

\section{results} \label{sec:result}

\subsection{Model Comparison} \label{Model comparison}

This section examines the properties in the different models introduced in Section~\ref{sec:setup}. Our 2D simulations comprise five distinct runs, each with a different initial magnetic field configuration. 

Adopting standard definitions, we define the mass accretion rate $\dot{M}$ and the magnetic flux threading the black hole horizon, $\Phi_{\rm BH}$, as follows \citep[e.g.,][]{2019ApJS..243...26P}:

\begin{equation}
\dot{M} = \int_0^{2\pi} \int_0^{\pi} \rho u^r \sqrt{-g} \, d\theta \, d\phi,
\end{equation}
\begin{equation}
\Phi_{\rm BH} = \frac{1}{2} \int_0^{2\pi} \int_0^{\pi} |B^r| \sqrt{-g} \, d\theta \, d\phi,
\end{equation}
where, $\rho$ is the rest-mass density, $u^r$ is the radial component of the four-velocities, $B^r$ is the radial component of the magnetic field, and $\sqrt{-g}$ denotes the square root of the metric determinant.

\begin{figure}
    \centering
    \includegraphics[width=1.01\columnwidth]{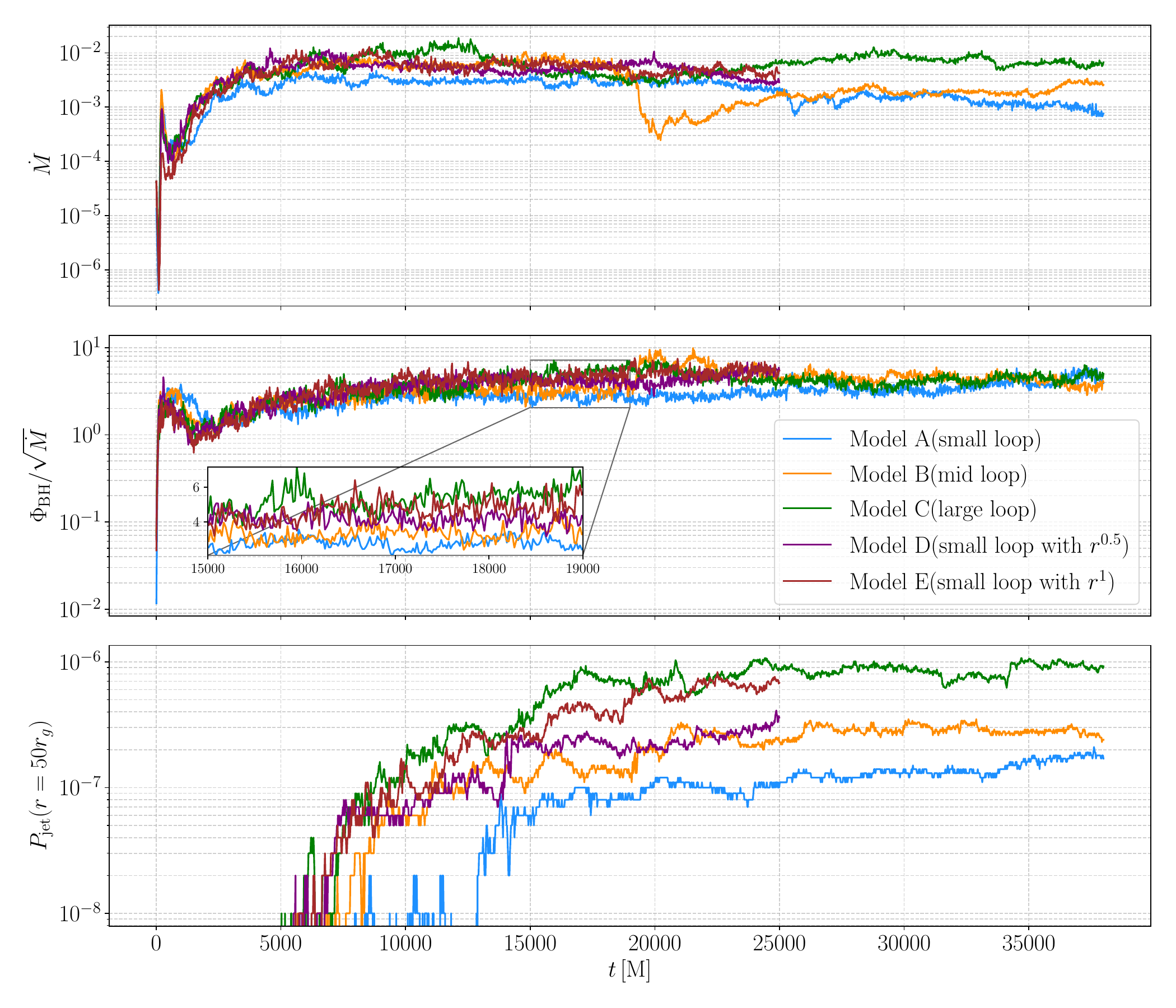}
    \caption{Time evolution of the mass accretion rate measured at the event horizon (top), the normalized magnetic flux at the horizon (bottom), and the jet power evaluated at $r = 50\,r_g$. The inset panel displays the normalized magnetic flux on a linear scale.
    } 
    \label{fig:MdotBfluxJetpow}
\end{figure}

As shown in Figure~\ref{fig:MdotBfluxJetpow}, the different models do not exhibit significant variations in the mass accretion rate $\dot{M}$ and the magnetic flux $\Phi_{\rm BH}$. This behavior contrasts with the magnetically arrested disk (MAD) scenario found in geometrically thick torus simulations by \cite{2023MNRAS.522.2307J}, 
which exhibits QPOs in both \(\dot{M}\) and \(\Phi_{\rm BH}\) due to the cyclical accumulation and eruptions of magnetic flux near the black hole horizon. 
In our case, the initially prescribed multi-loop, SANE-like magnetic field configuration prevents the buildup of large-scale magnetic fields at the horizon, thereby suppressing to enter MAD regime.
In all models, the normalized magnetic flux remains below the canonical threshold $\Phi_{\rm BH} / \sqrt{\dot{M}} \approx 15$, which is commonly adopted as the condition for the onset of the MAD regime \citep{2011MNRAS.418L..79T}. Furthermore, recent studies suggest that the MAD state may be incompatible with explaining the luminous hard state \citep{2023MNRAS.525L..82F}.

To quantify the differences in jet production between our models, we measure the jet power at $r = 50\,r_g$, following the definition of equation~(\ref{eq:Pjet}), 
\begin{equation}
P_{\rm jet} := \int_0^{2\pi} \int_0^{\pi} \left( -T^r_t - \rho u^r \right) \sqrt{-g} \, d\theta d\phi.
\label{eq:Pjet}
\end{equation}
The integrand in equation~(\ref{eq:Pjet}) is set to zero in the regions where the magnetization $\sigma \equiv b^2 / \rho \leq 1$, so that only the magnetically dominated outflows are included in the jet power.
In models with larger magnetic loops (Models~B and C), we observe enhanced jet power. This is primarily a consequence of the relatively higher magnetic flux $\Phi_{\rm BH}$ in these cases, as illustrated in the zoom-in panels. The difference in jet power is further amplified by the fact that, under the Blandford–Znajek mechanism, the jet power scales as $P_{\rm jet}^{\rm BZ} \sim \kappa \left( \frac{\Phi_{\rm BH}}{4\pi M} \right)^2 \Omega_{\rm H}^2$ \citep{1977MNRAS.179..433B}, where $M$ is the black hole mass, and $\Omega_{\rm H}$ is the angular frequency of the black hole horizon. As a result, the jet power differs by approximately one order of magnitude between Model~A (small loop case) and Model~C (large loop case), with Model~C exhibiting a factor of $\sim 10$ higher jet power. For Models~D and E, the inclusion of a radial dependence in the initial magnetic field configuration leads to a stronger magnetic field in the inner regions. This, in turn, results in a higher normalized magnetic flux and enhanced jet power. For the BP-driven wind, we find that it is two to three orders of magnitude stronger than the BZ-driven jet, consistent with previous studies\citep{2021MNRAS.505.3596D,2022MNRAS.517.5032D,2025JCAP...01..152D}.

\begin{figure*}
    \centering
    \includegraphics[width=\linewidth]{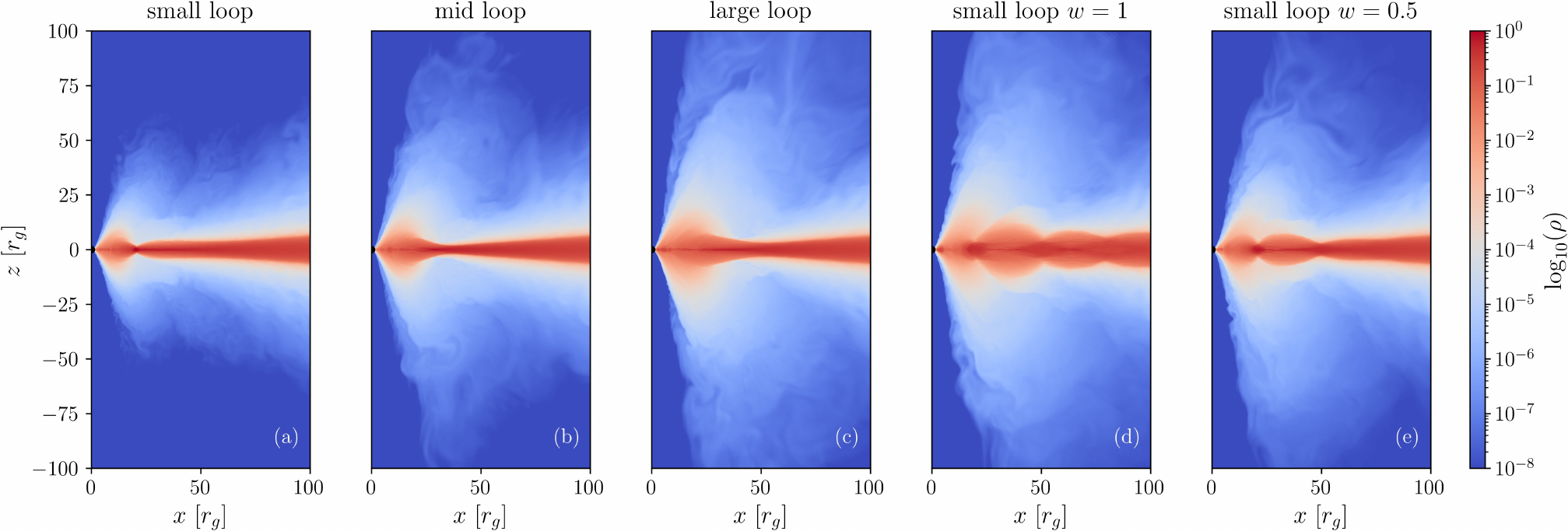}
    \caption{Logarithmic density distributions at $t=25{,}000\,\rm M$ for different magnetic configurations. The labels in the lower right corner of each panel indicate the corresponding model: (a) small loop, (b) mid loop, (c) large loop, (d) small loop with $A_\phi \propto r^1$, and (e) small loop with $A_\phi \propto r^{0.5}$. } 
    \label{fig:rho}
\end{figure*}

After the simulations reach a quasi-steady state, we analyze a snapshot at $t = 25{,}000\,\mathrm{M}$, as shown in Figure~\ref{fig:rho}. We find that in Figure~\ref{fig:rho}(c) (the large-loop case), the inner region expands more significantly in the vertical direction, extending up to $\sim 100\,r_g$. In contrast, the vertical extent is reduced in the medium- and small-loop cases, reaching $\sim 75\,r_g$ in Model~B and $\sim 50\,r_g$ in Model~C, respectively. For Figures~\ref{fig:rho}(d) and \ref{fig:rho}(e), where the magnetic field is strengthened by the inclusion of an additional radial dependence, we also observe a broader vertical expansion compared to the small-loop case. We clearly see multiple expanded loop-like magnetic structures along the equatorial plane, indicating that the initial magnetic loops continue to influence the magnetic field morphology even at late evolutionary stages.

\begin{figure*}
    \centering
    \includegraphics[width=\linewidth]{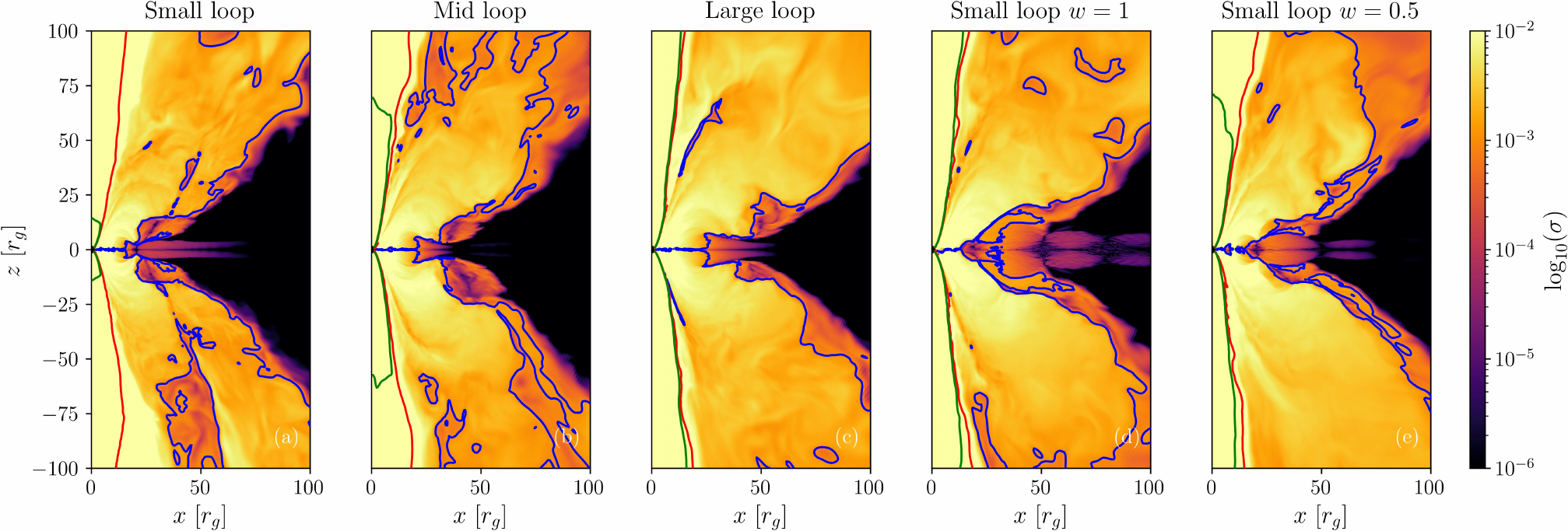}
    \caption{The figure shows the distribution of time-averaged magnetization $\sigma$ over the interval $t = 24{,}000\,\mathrm{M}$ to $25{,}000\,\mathrm{M}$. The labels in the figure correspond to the same model designations as those in Figure~\ref{fig:rho}. The blue, red, and green contour lines correspond to $\sigma = 10^{-3}$, $1$, and $10$, respectively.} 
    \label{fig:sigma}
\end{figure*}

We compute the time-averaged magnetization $\sigma$ for all five models over the interval $t = 24{,}000\,\mathrm{M}$ to $25{,}000\,\mathrm{M}$, as shown in Figure~\ref{fig:sigma}. As shown in Figures~\ref{fig:sigma}(a), (b), and (c), the region enclosed by the $\sigma = 10$ (green contour) becomes smaller when the loop size is smaller, indicating a reduction in jet power in the small-loop case. In contrast, Figures~\ref{fig:sigma}(d) and (e) show enhanced jet power compared to panel \ref{fig:sigma}(a), which can be attributed to the inclusion of radial dependence in the initial magnetic field configurations. This is consistent with the conclusion drawn from Figure~\ref{fig:MdotBfluxJetpow}, further supporting the dependence of both jet power and magnetic flux on the initial magnetic field configuration. We also find that panel~\ref{fig:sigma}(c) exhibits a higher $\sigma$ distribution with a broader vertical expansion in the inner region, compared to smaller loop cases of panels~\ref{fig:sigma}(a) and \ref{fig:sigma}(b). A similar trend is also observed in panels~\ref{fig:sigma}(d) and \ref{fig:sigma}(e). 

All models consistently exhibit a distinct two-region structure: a relatively thick, low-density inner region and a comparatively thin, high-density disk extending outward. To quantitatively characterize the separation between these two components, we introduce a density-weighted scale height following \citet{Noble2009}. The density-weighted shell average of any quantity $f$ is defined as
\begin{equation}
\langle f \rangle = \frac{\int_{\sigma < 1} f \rho \sqrt{-g} \, d\theta d\phi}{\int_{\sigma < 1} \rho \sqrt{-g} \, d\theta d\phi}, 
\label{eq:rhoavg}
\end{equation}
where the integration is performed over the gas-dominated region $\sigma < 1$. The scale height $H$ is then computed as the variance of the vertical density moment:
\begin{equation}
H \equiv \left( \langle z^2 \rangle - \langle z \rangle^2 \right)^{1/2}, 
\label{eq:scaleheight}
\end{equation}
where $z = r \cos \theta$ denotes the vertical coordinate, and both $\langle z \rangle$ and $\langle z^2 \rangle$ are evaluated using Eq.~\eqref{eq:rhoavg}. 

\begin{figure}
    \centering
    \includegraphics[width=0.8\columnwidth]{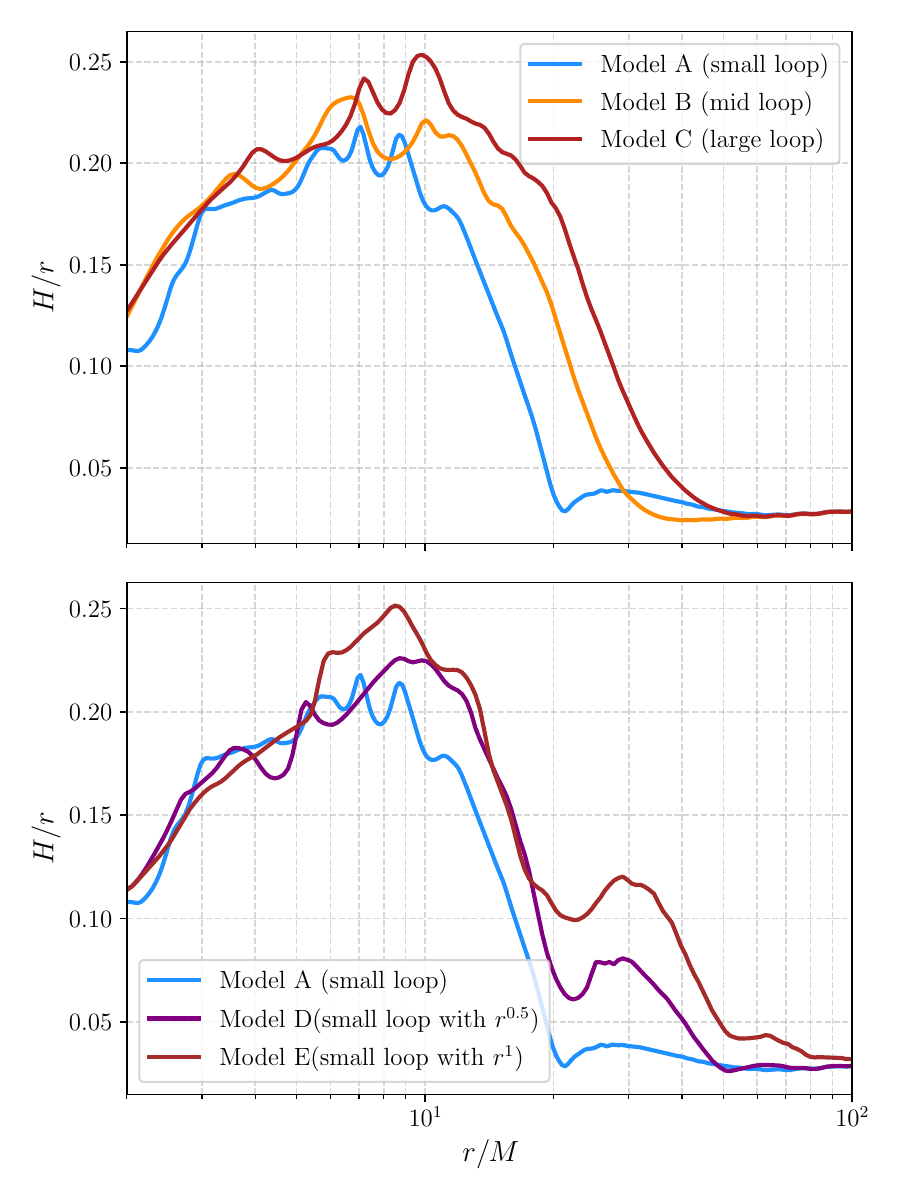}
\caption{Scale height profiles for different models, time-averaged over the interval $t = 24{,}000\,\mathrm{M}$ to $25{,}000\,\mathrm{M}$.}
    \label{fig:scale-height}
\end{figure}

Figure~\ref{fig:scale-height} shows the time-averaged scale height for different models over the interval $t = 24{,}000\,\mathrm{M}$ to $25{,}000\,\mathrm{M}$. We observe that the inner region exhibits a higher scale height, corresponding to a broader vertical expansion, while the outer region maintains a relatively low scale height of $\sim 0.03$, consistent with the typical criterion for thin disks adopted in several previous studies \citep{2022ApJ...935L...1L,2025MNRAS.541.3184H}. As shown in the upper panel of Figure~\ref{fig:scale-height}, the large-loop case reaches the thin-disk region at a larger radius, which we refer to as the truncation radius. Model~A exhibits a truncation radius of approximately $60\,r_g$, while Model~B has a truncation radius of $\sim 40\,r_g$. For Model~C, the first noticeable truncation occurs around $20\,r_g$. As the trend in Figure~\ref{fig:scale-height} indicates, a larger initial magnetic loop size corresponds to an increase in the truncation radius. We notice that the blue curve corresponding to the small-loop case exhibits a secondary rise in scale height around $r \sim 30\,r_g$. This feature is further amplified in the bottom panel, where models with radial components in the magnetic field configuration show a more pronounced secondary rise in scale height. In particular, Models~D and E exhibit this second rise more clearly. By combining the scale height profiles in Figure~\ref{fig:scale-height} with the density profile shown in Figure~\ref{fig:rho}, we find that the vertical expansion is strongly dependent on the initial magnetic loop configuration.

\subsection{Truncation Radius and Puff-Up Region}

The structure of the inner accretion flow exhibits significant variation depending on the initial magnetic field configuration. In particular, both the location at which the flow transitions into a thin disk, the truncation radius, and the degree of vertical expansion in the inner region are found to be sensitive to the magnetic loop topology. 

Models with larger magnetic loops or enhanced radial dependence tend to produce more extended geometrically thick regions near the black hole, as well as larger truncation radii. This correlation suggests that the initial field configuration not only influences jet launching efficiency but also plays a critical role in shaping the vertical structure of the inner accretion flow.

\begin{figure*}
    \centering
    \includegraphics[width=\textwidth]{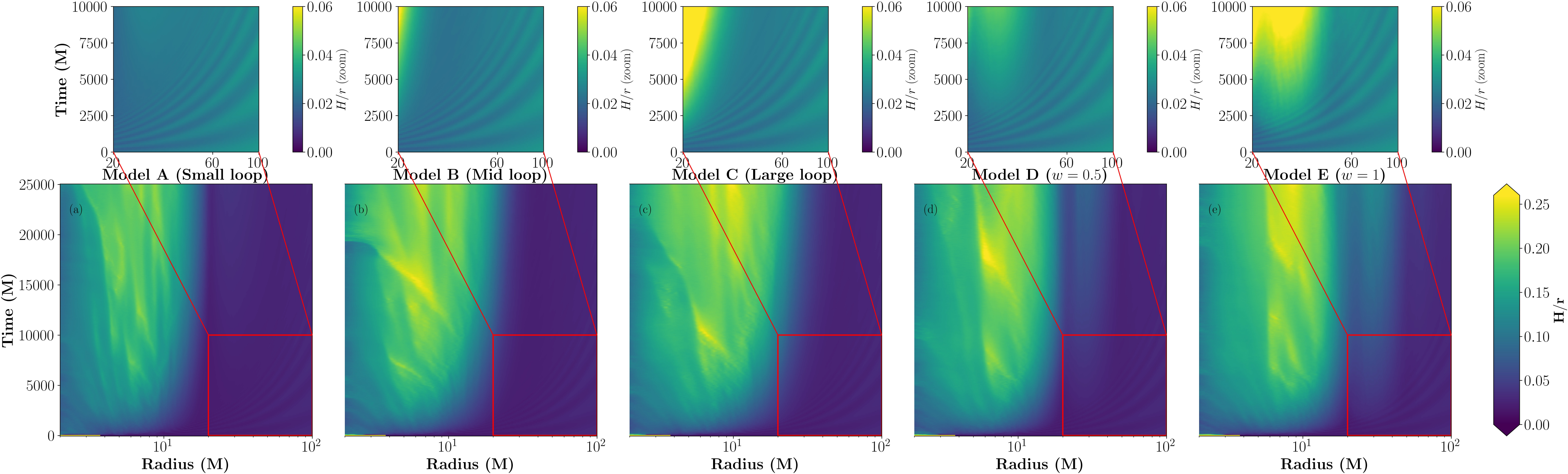}
    \caption{
Space-time distribution of the scale height for different models. The labels in the upper-left corner of each panel indicate the corresponding model: (a) small loop, (b) mid loop, (c) large loop, (d) small loop with $A_\phi \propto r^1$, and (e) small loop with $A_\phi \propto r^{0.5}$. Note that the zoomed-in regions use different color bar ranges.
}
    \label{fig:timespace-scaleheight}
\end{figure*}

In Figure~\ref{fig:timespace-scaleheight}, we present the space–time diagram of the scale height. A comparison among Models~A, B, and C clearly shows that larger magnetic loops lead to a higher scale height in the inner region, i.e., more pronounced vertical expansion. In Models~D and E, the inclusion of a radial dependence in the magnetic field configuration further amplifies the scale height, and in Model~E, a secondary expansion feature is observed at larger radii. We also observe inclined stripe-like patterns in the zoomed-in panels across all models, suggesting coherent features in the time–radius evolution of the scale height. A more detailed explanation of these features will be provided in Section~\ref{beta}.

\subsubsection{Plasma $\beta$} \label{beta}

To further investigate the physical origin of the vertical expansion observed in the inner region of our simulated models, we examine the distribution of plasma beta ($\beta$), defined as the ratio of gas pressure to magnetic pressure, $\beta = p_{\rm gas}/p_{\rm mag}$.
This dimensionless parameter indicates the dominant pressure component in the plasma: regions with $\beta < 1$ are magnetically dominated, whereas $\beta > 1$ implies thermal pressure dominance. Previous studies (e.g., \citealt{2025JCAP...01..152D}) have shown that the appearance of low-$\beta$ regions in the inner disk can lead to vertical expansion due to enhanced magnetic pressure support and magnetic buoyancy. 

\begin{figure*}
    \centering
    \includegraphics[width=\textwidth]{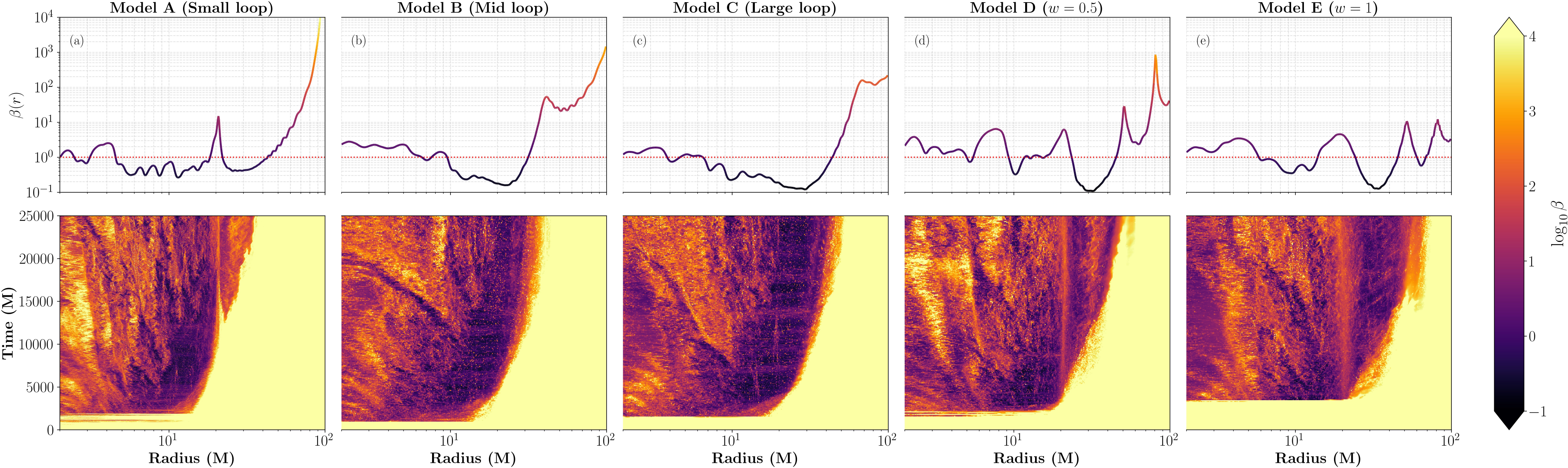}
    \caption{
Upper panels: time-averaged plasma $\beta$ over the interval $t = 24{,}000\,\mathrm{M}$ to $25{,}000\,\mathrm{M}$, computed using shell-integrated values. Bottom panels: space–time diagrams of plasma $\beta$, illustrating the temporal evolution of magnetization at different radii.
}
    \label{fig:beta}
\end{figure*}

In Figure~\ref{fig:beta}, we present the time and spatial evolution of the plasma $\beta$. From the upper panels, the smaller-loop models exhibit spatially confined regions of $\beta < 1$, indicating that magnetic pressure dominates over gas pressure in more confined zones. Model~C, with the large-loop configuration, exhibits a broad region where $\beta$ extends from approximately $4\,r_g$ to $45\,r_g$. In comparison, Model~B shows a narrower low-$\beta$ region, ranging from $\sim 10\,r_g$ to $30\,r_g$. The small-loop case exhibits two distinct regions where $\beta < 1$, both corresponding to the initial magnetic loop configuration. The first low-$\beta$ region extends from approximately $6\,r_g$ to $17\,r_g$, while the second spans from $\sim 23\,r_g$ to $48\,r_g$.

From the bottom panels, we observe that the low-$\beta$ regions tend to extend outward in radius over time. In Model~A, this radial extension stabilizes around $t = 10{,}000\,\mathrm{M}$, after which the second magnetic loop begins to dominate the evolution at larger radii. Similarly, the low-$\beta$ regions in Models~B and C stabilize at later times, around $t = 15{,}000\,\mathrm{M}$ and $t = 25{,}000\,\mathrm{M}$, respectively. 
For Models~D and E, the upper panels show that low-$\beta$ regions extend to larger radii compared to the small-loop case. In particular, Model~E exhibits a third distinct region where $\beta < 1$, indicating the presence of magnetically dominated zones at even larger distances from the black hole.

By comparing the high scale-height regions in Figure~\ref{fig:scale-height} with the low-$\beta$ regions in Figure~\ref{fig:beta}, we identify a strong spatial correlation between the two quantities. This suggests that the enhanced magnetic pressure in low-$\beta$ regions may be one of the primary contributors to the vertical expansion observed in the inner disk.

However, we find that in the early stages of the evolution, all models exhibit regions of very high plasma $\beta$. This is inconsistent with the scale height profiles shown in Figure~\ref{fig:scale-height}, where vertical expansion begins at early times across all models. This discrepancy suggests that plasma $\beta$ alone may not fully account for the onset of vertical expansion. Additionally, in Model~E, we observe a region of elevated scale height between $r \sim 30\,r_g$ and $t = 6000$–$10{,}000\,\mathrm{M}$ in Figure~\ref{fig:scale-height}, which does not correspond to any clear feature in the plasma $\beta$ distribution shown in Figure~\ref{fig:beta}. Moreover, the plasma $\beta$ profile alone cannot account for the appearance of the inclined, stripe-like structures observed in the zoomed-in panels of Figure~\ref{fig:scale-height}. Taken together, these inconsistencies motivate the search for an alternative physical mechanism that can account for the additional structures observed in the scale height profiles.

\subsubsection{ $\alpha$ viscosity } \label{alpha}

The discrepancies between the plasma $\beta$ distribution and the early onset of vertical expansion, as well as the appearance of inclined stripe-like structures, suggest that additional mechanisms may be at play. One plausible candidate is the turbulent angular momentum transport associated with an effective viscosity driven by MRI in the disk. It is often parametrized using the dimensionless $\alpha$-viscosity model.

Following \citet{2024MNRAS.533..254S}, we adopt the Maxwell $\alpha$ prescription defined by
\begin{equation}
\alpha_M \equiv -\frac{\left\langle b^r \sqrt{g_{rr}} b^\phi \sqrt{g_{\phi\phi}} \right\rangle^{\rm disk}_{\rho[\theta,\phi]}}{\left\langle p_t \right\rangle^{\rm disk}_{\rho[\theta,\phi]}}.,
\end{equation}
where \( p_t = b^2 / 2 + p_g \)  is the total pressure, and the brackets follow the same definition as Eq.~\eqref{eq:rhoavg}. This formulation captures the strength of local turbulent transport and its relation to the gas pressure, allowing us to quantify angular momentum redistribution within the disk.

\begin{figure*}
    \centering
    \includegraphics[width=1.08\textwidth]{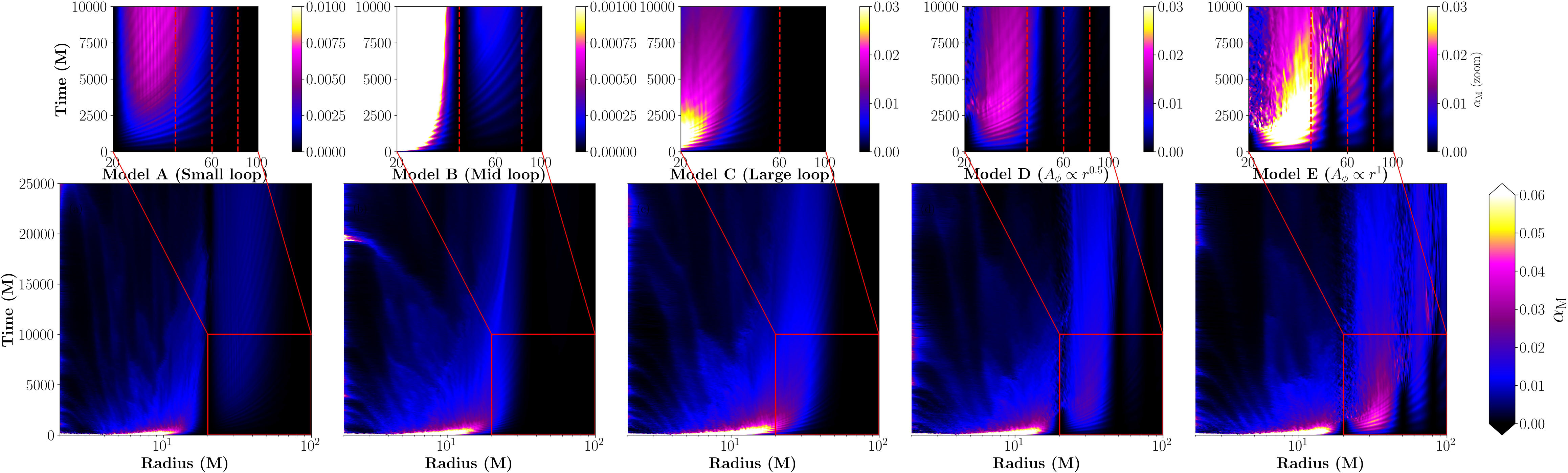}
    \caption{
Same labeling as in Figure~\ref{fig:timespace-scaleheight}, but shown for $\alpha_M$ in different models. Red dash lines refer to the interfaces between loops.
}
    \label{fig:spacetime-alpha}
\end{figure*}

Figure~\ref{fig:spacetime-alpha} presents the space-time evolution of $\alpha_M$.
In the zoomed-in region, distinct inclined stripe-like features can be seen, which closely correlate with the morphology of the local scale height shown in Figure~\ref{fig:timespace-scaleheight}.
These patterns are consistent with radially propagating oscillations in a viscous, compressible disk.
In such flows, small perturbations in the pressure and stress fields naturally excite inertial-acoustic waves that travel through regions of varying sound speed and thickness, forming coherent slanted streaks in time-radius space \citep{2003MNRAS.344L..37R}.
The strong correlation between $\alpha_M$ and the scale height indicates that these oscillations involve coupled variations of stress, pressure, and vertical support, consistent with the hydrodynamic feedback between viscosity and disk thickness \citep{2012MNRAS.426.1107L}.

The outer, geometrically thinner portion of the disk exhibits the most coherent stripe-like features, implying longer-lived, radially organized oscillations.
In contrast, the inner, thicker region lacks such ordered patterns: enhanced turbulence and stronger viscous diffusion there damp the radial coherence of the oscillations, producing more stochastic, short-lived fluctuations.
This transition—from coherent wave-like patterns in the thin outer disk to irregular variability in the thick inner flow—suggests that the ability of the disk to sustain organized oscillations diminishes as its geometric thickness increases \citep{2009MNRAS.393..979L}.

\subsubsection{Time variability and QPO-like features} \label{time-var}

\begin{figure}
    \centering
    \includegraphics[width=1\linewidth]{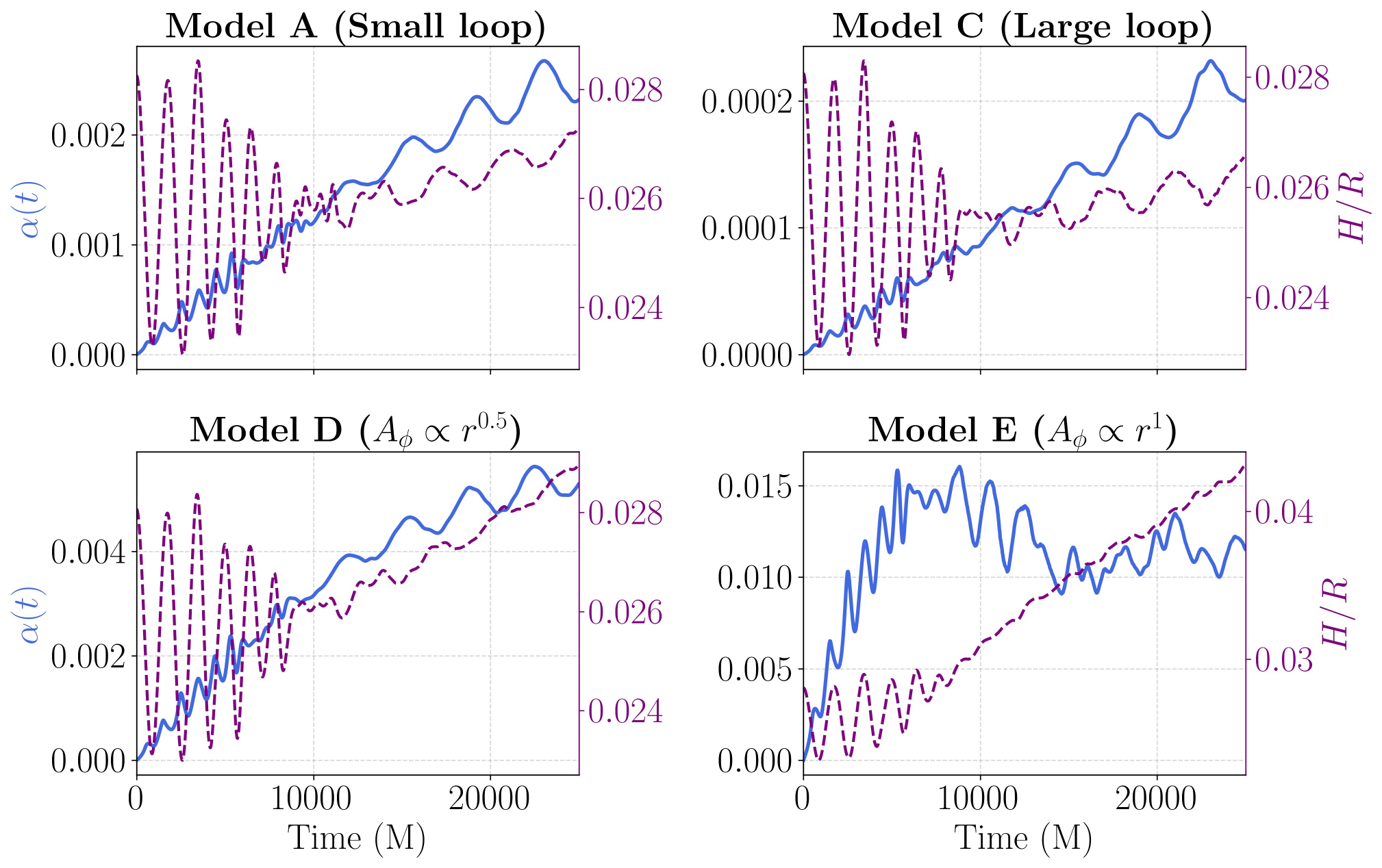}
\caption{Time evolution of $\alpha_M$ and the scale height at $r = 60\,r_g$ for four simulation models.}
    \label{fig:r60}
\end{figure}

To more clearly illustrate the temporal periodicity of these oscillations, we present the time series of the $\alpha_M$ viscosity at selected radii.
Figure~\ref{fig:r60} shows the evolution of $\alpha_M$ and the scale height at $r = 60\,r_g$ for the four models.
During the early stage ($t \lesssim 10^4\,\rm M$), all four models display clear periodic oscillations, consistent with the quasi-periodic, inclined stripe-like structures seen in the space-time diagrams.
Both Models~A and C show that the scale height stabilizes around $H/R \sim 0.03$ after the initial oscillations, with Model~A exhibiting a higher $\alpha_M$ amplitude—likely because smaller magnetic loops promote stronger secondary viscous fluctuations.
Introducing a radial dependence in Model~D enhances this secondary modulation, further increasing the effective stress amplitude.

In Models~A–C, both $\alpha_M$ and the disk scale-height exhibit coherent QPOs with comparable timescales, consistent with inertial–acoustic motions near the radial epicyclic frequency $\nu_r$ in the inner disk \citep{2012MNRAS.426.1107L}.
The physical origin of these oscillations can be understood within the viscous–epicyclic overstability framework, in which local stress perturbations interact nonlinearly with orbital shear and pressure variations \citep{2015MNRAS.446..240M}.
When the viscous response of the flow is phase-shifted relative to compressive motions, the shear stress performs positive work on the oscillation, leading to self-excited, growing inertial-acoustic modes at frequencies close to $\nu_r$.
The initial multi-loop magnetic field topology reinforces this process by creating a series of confined cavities that act as resonant traps for viscous–acoustic perturbations, allowing the overstability to operate locally and sustain long-lived modulation of $\alpha_M$ and disk thickness. 
Regions with steep radial gradients in $\nu_r$ enhance overstability by the coupling between radial and vertical motions, thereby promoting efficient energy exchange between viscous and compressive modes.
Conversely, in geometrically thicker regions where the restoring epicyclic force weakens and diffusion dominates, the feedback becomes inefficient and the oscillations lose coherence.
During the early simulation time, this viscous–epicyclic overstability naturally maintains the quasi-periodic modulation of $\alpha_M$ and $H/R$, with characteristic frequencies determined by local epicyclic dynamics. 

In contrast, Model~D undergoes a rapid geometric thickening of the inner flow after the initial oscillatory phase.
As the disk inflates, the radial epicyclic restoring force weakens and the effective viscous diffusivity increases, which disrupts the phase relation required for viscous–epicyclic overstability.
The self-excited inertial-acoustic oscillations that dominated the early, thin-disk stage are therefore gradually damped as the overstability is quenched by enhanced diffusion and reduced epicyclic confinement.
This behavior is consistent with linear analyses showing that viscous overstability is most effective in geometrically thin disks with strong $\nu(r)$ gradients but becomes suppressed once the disk is thicker and the acoustic trapping condition is lost \citep{2015MNRAS.446..240M}.
Together with the increase in plasma $\beta$ shown in Figure~\ref{fig:beta}, this transition marks a change in the dominant vertical support—from $\alpha_M$-modulated inertial oscillations to magnetically dominated pressure support when the disk evolves into its thick, turbulent state.

\begin{figure}
    \centering
    \includegraphics[width=1\linewidth]{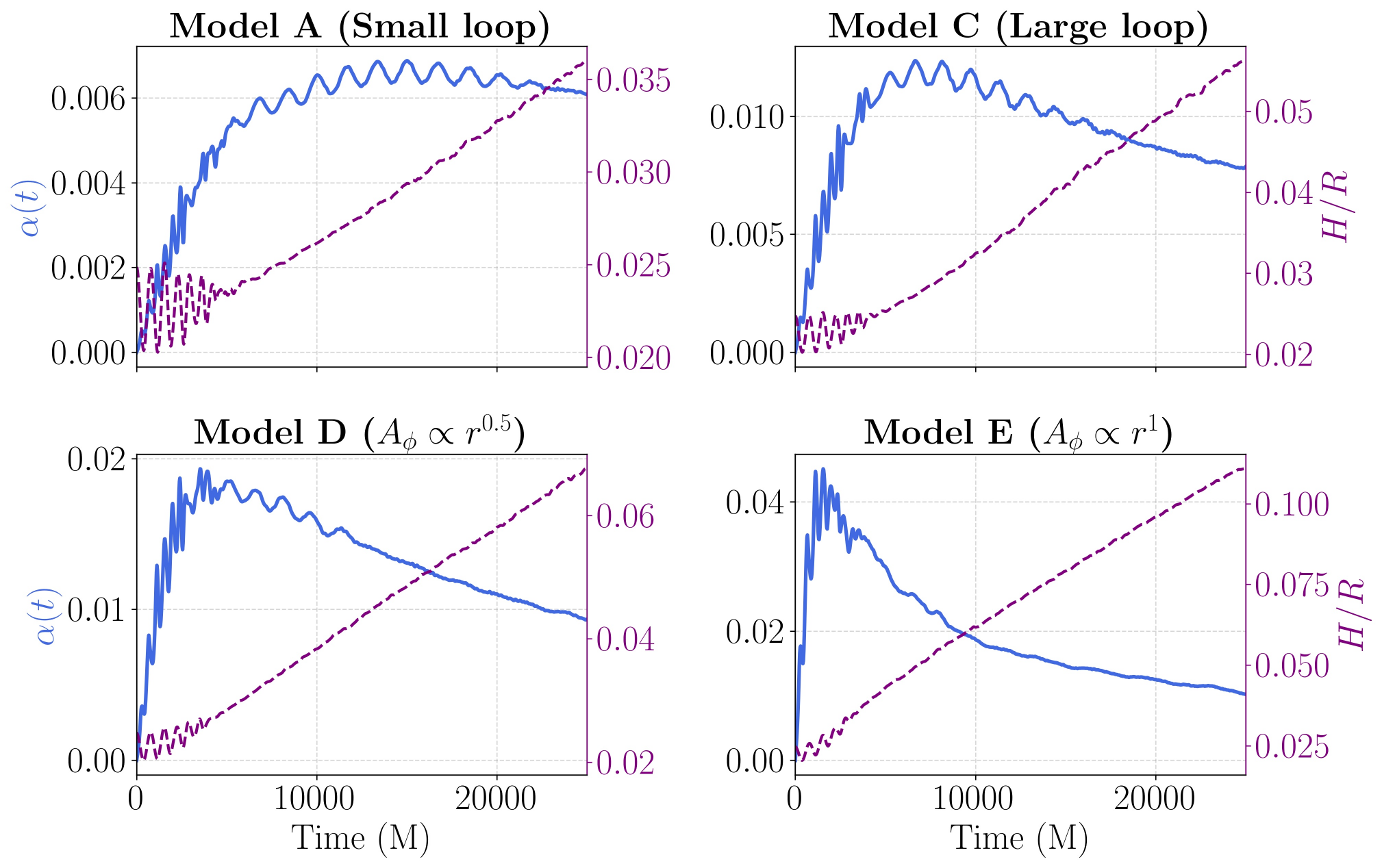}
\caption{Same as Figure~\ref{fig:r60} but at $r = 35\,r_g$}
    \label{fig:r35}
\end{figure}

Figure~\ref{fig:r35} presents the time evolution of $\alpha_M$ at $r = 35\,r_g$, illustrating how the overstable behavior manifests itself in the inner flow.
At this smaller radius, both $\alpha_M$ and the scale height show coherent high-frequency modulations that mirror the stripe-like features seen in the spacetime diagrams.
These fluctuations correspond to viscous overstable inertial-acoustic oscillations tuned to the local epicyclic frequency $\nu_r$, where neighboring fluid rings execute self-excited radial motions about their equilibrium orbits.
The nested magnetic loops formed in the initial state provide partial boundaries that confine these oscillations and sustain their growth for several dynamical times.

When the plasma $\beta$ increases and the disk vertically expands, it erases the trapping cavities that supported the overstability, leading to a gradual loss of coherence and the disappearance of the stripe patterns.
Among the cases, Model~A maintains this activity for the longest duration, followed by Models~C and D, while Model~E shows only transient oscillations.
This trend confirms that the persistence of viscous overstability depends jointly on disk thickness and magnetic topology. Thin disks with strong epicyclic gradients favor sustained coherent modes, whereas thicker, more diffusive flows rapidly damp them through viscous diffusion.

\begin{figure}
    \centering
    \includegraphics[width=0.85\linewidth]{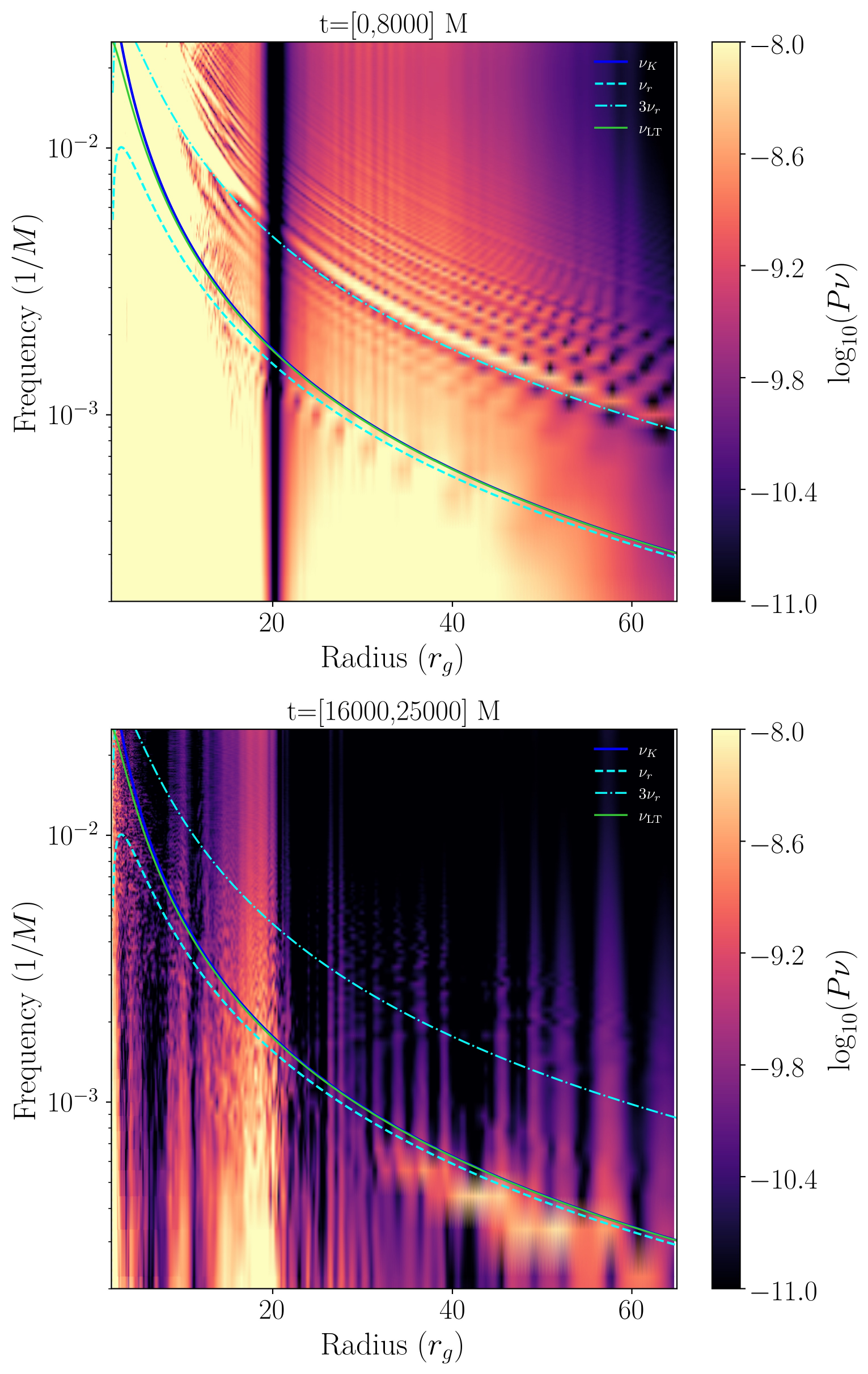}
\caption{PSDs of $\alpha_M$ measured within the inner $65\,r_g$ of the accretion disk in Model~A (small-loop). 
The upper panel is computed over the first $8000\,\mathrm{M}$ time period, while the lower panel corresponds to the later phase $t=[16000,\,25000]\,\mathrm{M}$. 
The dashed cyan, solid blue, and solid green lines indicate the radial epicyclic, Keplerian, and Lense-Thirring precession frequencies, respectively, while the dash-dotted cyan line marks three times the radial epicyclic frequency. 
The color scale represents $\log(P\cdot\,\nu)$, corresponding to the logarithmic spectral power.}
    \label{fig:PSD-2D}
\end{figure}

To further probe the dynamical behavior in this region, we analyze the temporal variability of the accretion flow by computing the power spectral density (PSD) of the viscosity $\alpha_M$ at different radii. 
The PSD is obtained as
\begin{equation}
P(\nu) = |\tilde{\alpha}_M(\nu)|^2,
\end{equation}
where $\tilde{\alpha}_M(\nu)$ is the Fourier transform of the time-series of $\alpha_M$ within a given radial shell. 
In practice, we apply the discrete Fourier transform, and the resulting PSDs are proportional to the variance of the signal. Figure~\ref{fig:PSD-2D} shows the PSDs of Model~A over the time intervals $t = 0$--$8000\,\mathrm{M}$ and $t = 16000$--$25000\,\mathrm{M}$, overplotted with the characteristic frequencies  \citep{1998tbha.conf..233N}. The Keplerian frequency (solid blue curve) is given by
\begin{equation}
\nu_{k} = \frac{1}{2\pi}\left(r^{3/2} + a\right)^{-1},
\end{equation}
the radial epicyclic frequency (dashed cyan curve) by
\begin{equation}
\nu_{r}^{2} = \nu_{k}^{2}\left(1 - \frac{6}{r} + \frac{8a}{r^{3/2}} - \frac{3a^{2}}{r^{2}}\right),
\end{equation}
and the Lense–Thirring precession frequency (solid green curve) by
\begin{equation}
\nu_{\mathrm{LT}} = \frac{1}{2\pi}\,\frac{2a}{r^{3}},
\end{equation}
where $a$ is the black hole spin and $r$ is the radius. Throughout this work, we adopt geometrized units with $G = c = 1$.

At early times in the evolution (the first $8000\,\mathrm{M}$), bright bands appear along the curves corresponding to $\nu_r$ and its integer multiples. This radial dependence of QPOs is consistent with the trend expected from Figures~\ref{fig:r60} and \ref{fig:r35}: within the same model, QPOs originating at larger radii exhibit lower oscillation frequencies. This pattern indicates that the coherent oscillations of $\alpha_M$ identified in the time domain correspond to the radial motions of fluid oscillating near their natural epicyclic frequencies.
In this picture, the early-time PSD ridges trace a family of viscously amplified radially confined inertial acoustic modes—analogous to the high-frequency QPOs (HFQPOs) discussed in \citet{2018MNRAS.474.3967D}.
In the later stages of the evolution, the disk becomes thicker as the plasma $\beta$ increases, enhancing viscous diffusion and damping the oscillatory component.
As shown in Figure~\ref{fig:PSD-2D}, the overall PSD power decreases significantly, and the bright bands revert to the low-frequency Keplerian-driven QPO behavior.
The high-frequency signatures of the $\alpha_M$ oscillations associated with multiple $\nu_r$ harmonics vanish, consistent with the behavior previously discussed in Figures~\ref{fig:r60} and \ref{fig:r35}.

\subsection{3D Case} \label{3D}

\begin{figure}
    \centering
    \includegraphics[width=0.9\linewidth]{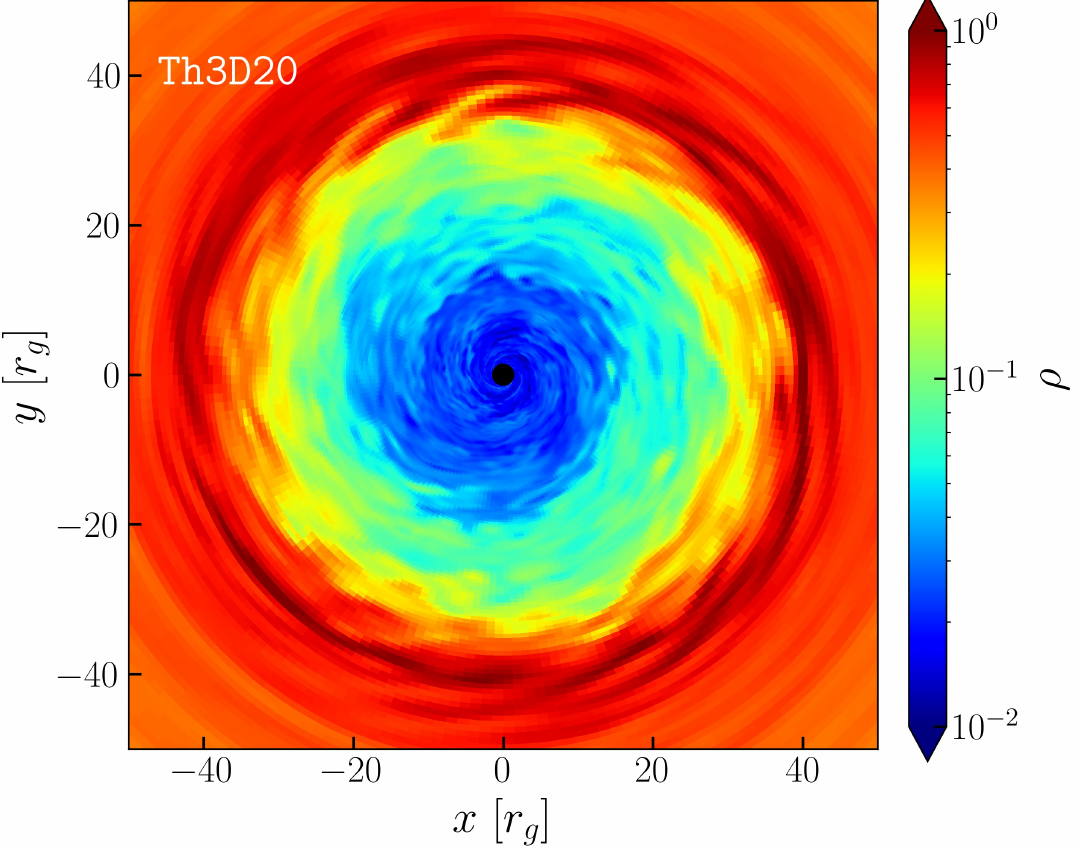}
\caption{Horizontal slice of logarithmic density profile at equatorial plane for \texttt{Th3D20} at $t=17900\,\rm M$.}
    \label{fig:rho-3D}
\end{figure}

To assess the robustness of the behavior observed in our 2D simulations, we additionally perform a fully three-dimensional simulation with identical initial conditions. As shown in Figure~\ref{fig:rho-3D}, the density snapshot at $t = 17900\, \rm M$ reveals a puffed-up, low-density region in the inner disk, closely resembling the structure observed in the 2D simulations. It shows a horizontal slice image at the equatorial plane.
Distinct transition features are observed at $ r \approx 20\,r_g$ and $ r \approx 40\,r_g$, indicating that our multi-loop magnetic configuration is effectively operating as intended. To ensure that our 3D simulations have sufficient resolution to resolve and follow the evolution of these structures, we discuss it in Appendix~\ref{appenC}.

\begin{figure}
    \centering
    \includegraphics[width=0.85\linewidth]{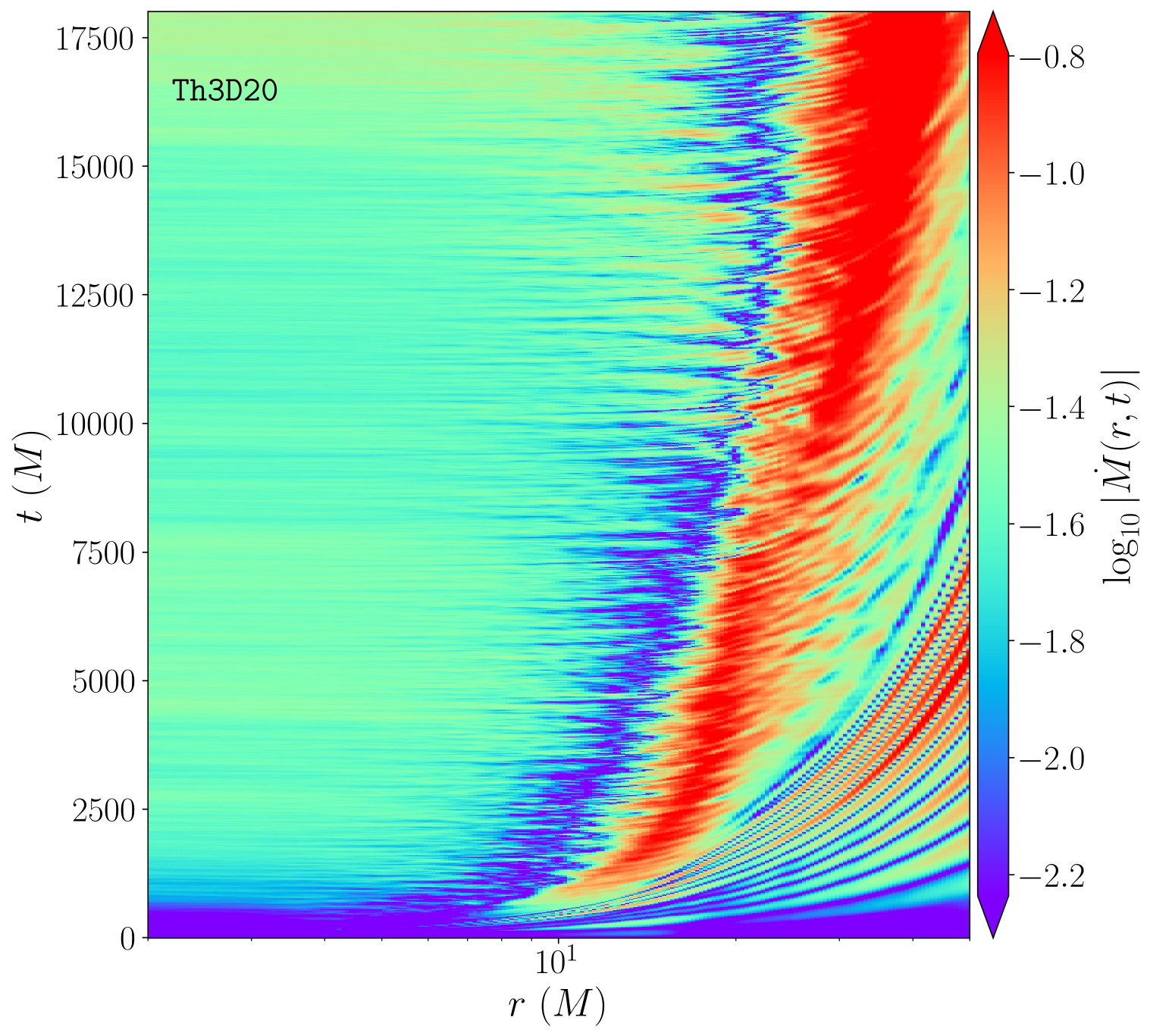}
\caption{Space-time diagram of mass accretion rate for model \texttt{Th3D20}.}
    \label{fig:3D-spacetime-Mdot}
\end{figure}

In Figure~\ref{fig:3D-spacetime-Mdot}, we present the spacetime diagram of the mass accretion rate $\dot{M}$ at different radius of model \texttt{Th3D20}.
Similar to the 2D simulations, we observe a series of inclined, stripe-like patterns at larger radii. 
As in the 2D case, we attribute these features to outward-propagating acoustic–inertial waves.
The inclined stripe-like patterns seen in the spacetime diagrams represent radially propagating inertial–acoustic perturbations sustained by viscous–epicyclic overstability. 
Their inclination corresponds to the finite group velocity of the acoustic waves, while their temporal spacing reflects the local epicyclic period. 
The coherence and persistence of these stripes indicate that the oscillations are global, self-excited modes rather than stochastic turbulence. 
The multi-loop magnetic topology provides partial reflection surfaces that confine these waves, allowing the stripes to appear as quasi-periodic, slanted bands across the disk.

\begin{figure}
    \centering
    \includegraphics[width=0.85\linewidth]{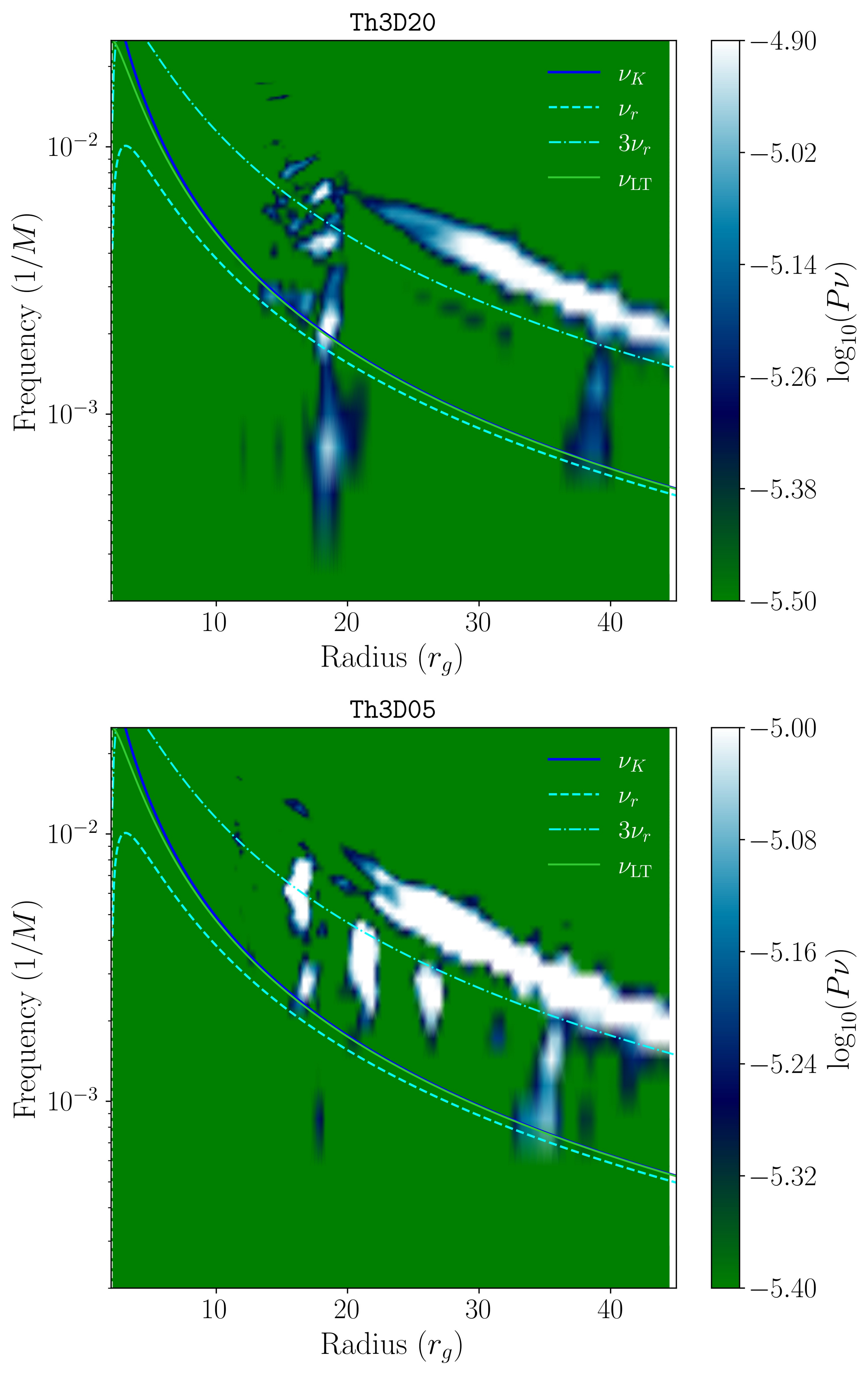}
\caption{PSD of the mass accretion rate \(\dot{M}\) for Models~ \texttt{Th3D20} (top) and \texttt{Th3D05} (bottom)
during the time period of $t=4000\,\rm M-8000\,\rm M$ . }
    \label{fig:PSD-3D}
\end{figure}

\begin{figure}
    \centering
    \includegraphics[width=0.85\linewidth]{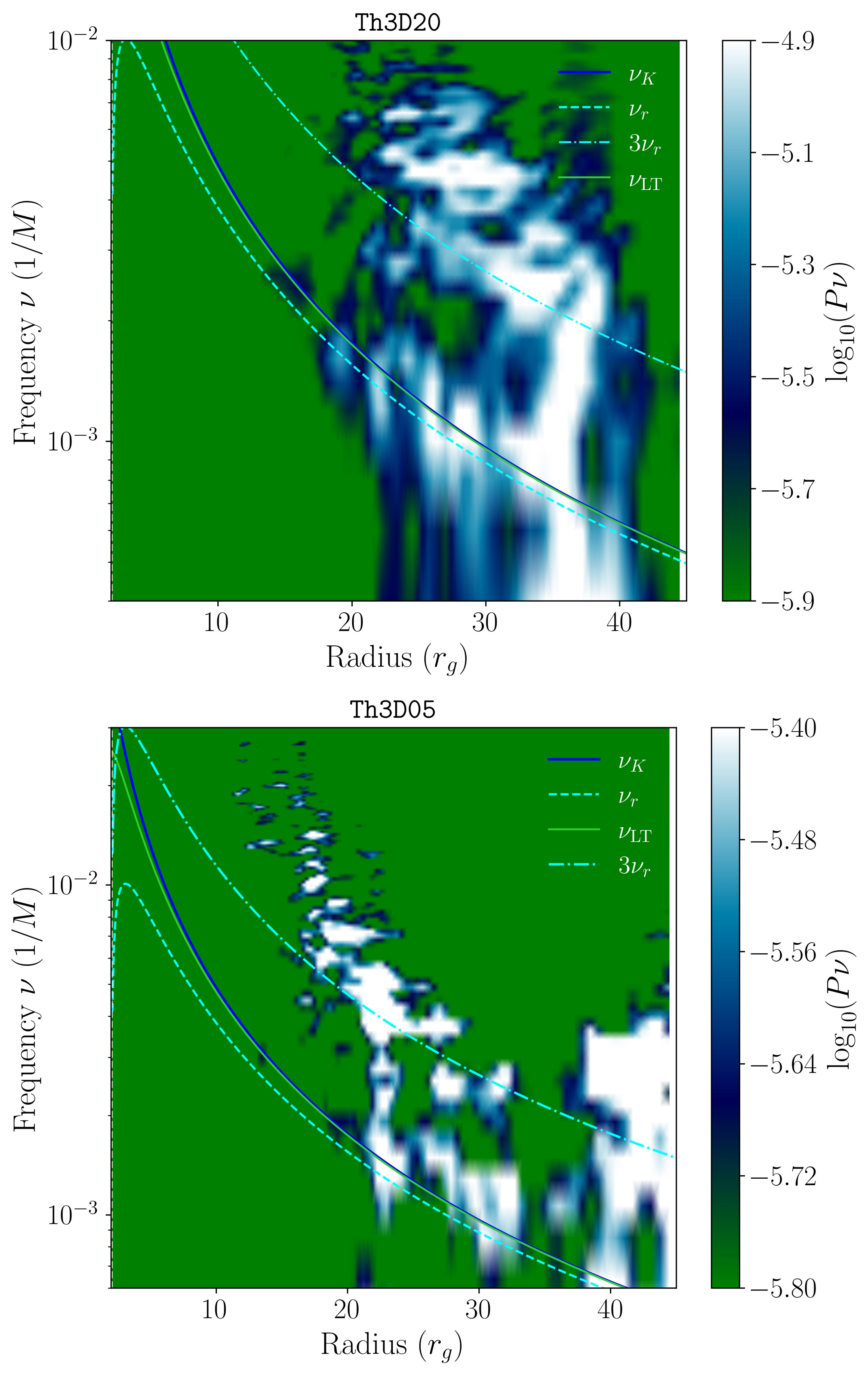}
\caption{Same as Figure~\ref{fig:PSD-3D}, but calculated during the time period of $t=12,000\,\rm M-18,000\,\rm M$ . }
    \label{fig:PSD-3D-late}
\end{figure}

At \(r = 20\,r_g\) and \(r = 40\,r_g\) in Figure~\ref{fig:3D-spacetime-Mdot}, we also see clear periodic variability in the mass accretion rate, \(\dot{M}\). 
These quasi-periodic features are consistently reflected in Figure~\ref{fig:PSD-3D}, where they appear as coherent, oscillatory patterns in time.
Similar to the 2D results, we observe a global bright band that follows the $3\,\nu_r(r)$ curve in Figure~\ref{fig:PSD-3D}, corresponding to the third harmonic of the viscously overstable inertial–acoustic mode discussed in Section~\ref{time-var}.
In addition, distinct bright bands appear near $r \approx 20\,r_g$ and $r \approx 40\,r_g$. These bands are associated with the size of the magnetic loop and are spatially coincident with the density transition layers identified in Figure~\ref{fig:rho-3D}. From the PSD of model~\texttt{Th3D05}, we also find a clear correlation between the loop size and the QPO location.
These boundary-associated QPOs originate from localized shear interfaces where sharp density (and $\alpha_M$ viscosity) gradients produce partial reflection of inertial–acoustic perturbations. At later times, as illustrated in Figure~\ref{fig:PSD-3D-late}, we see an enhancement of spectral power along the Keplerian frequency curve. This feature exhibits a transition consistent with our previous findings in 2D simulations. 

\begin{figure}
    \centering
    \includegraphics[width=0.85\linewidth]{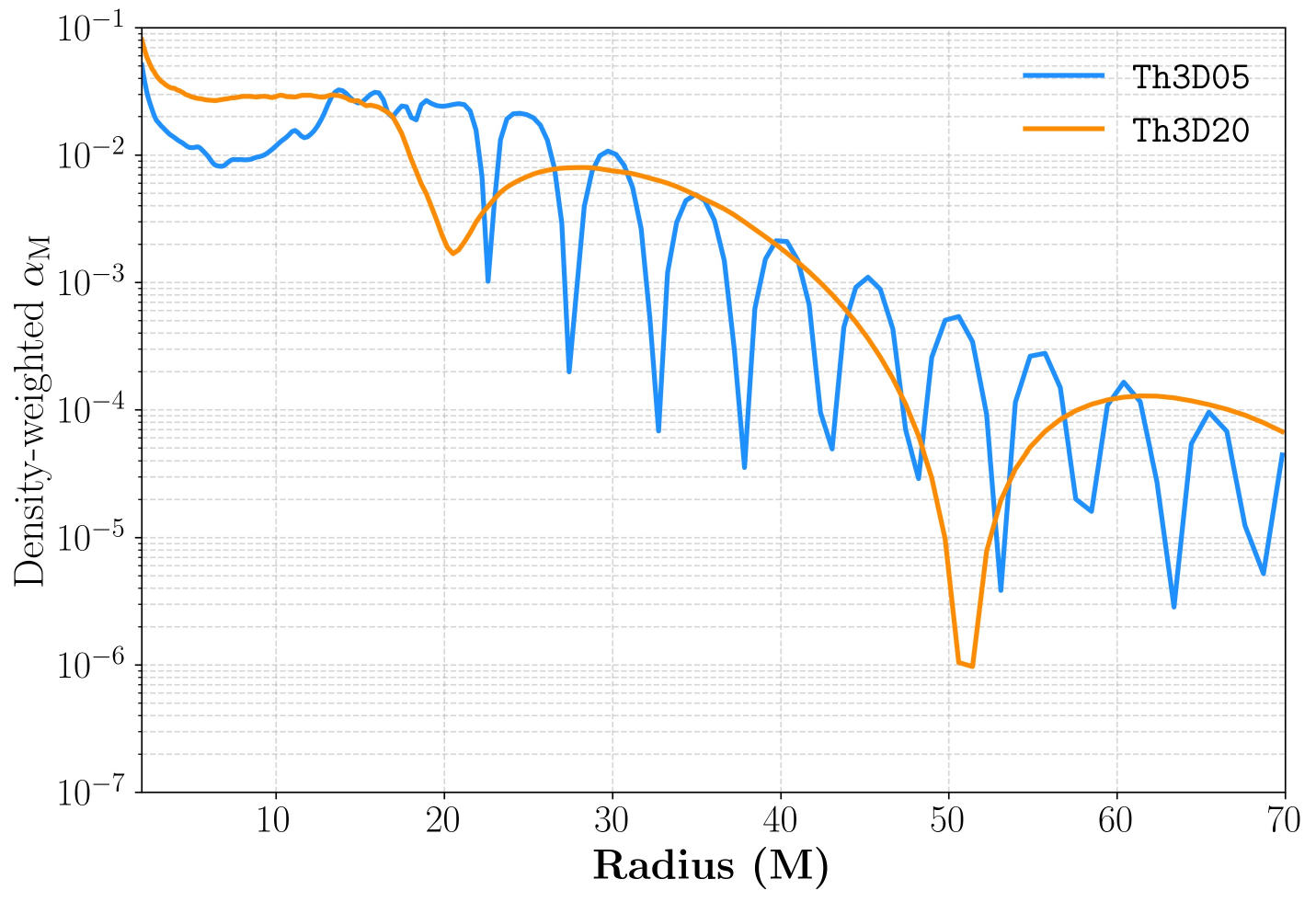}
\caption{The time-averaged and density-weighted $\alpha_{M}$ distribution calculated over the time period of $t = 9000\,\rm M$–$10,000\,\rm M$ of \texttt{Th3D20} (orange) and \texttt{Th3D05} (blue).}
    \label{fig:r-alpha}
\end{figure}

Figure~\ref{fig:r-alpha} presents the time-averaged and density-weighted $\alpha_{M}$ profile. The loop structure clearly produces a cavity (dip) of $\alpha_{M}$ which corresponds with the bright bands identified in the PSD maps (Figure~\ref{fig:PSD-3D}).
The resulting interface modes behave as small resonant cavities: perturbations trapped between the adjacent high- and low-density regions undergo repeated amplification through viscous feedback, giving rise to narrow, quasi-periodic power enhancements confined to the boundary zones.
Unlike the global $3\,\nu_r$ feature, these oscillations are spatially localized and governed by the local shear and density contrast rather than the global epicyclic structure. Further results of the loop size and the location of the interface are provided in the Appendix~\ref{appenD}.

In the viscous-epicyclic overstability framework, the key quantity that determines whether inertial-acoustic perturbations are damped or amplified is the relationship of phase between pressure and turbulent stress \citep{2022MNRAS.510..146H}. 
If the stress responds almost instantaneously to pressure fluctuations, it acts as a purely diffusive viscosity and does little
net work over an oscillation cycle. 
In contrast, a finite suitably signed lag between stress and pressure allows the turbulence to tap the free energy of differential rotation and feed it back into the wave, rendering the mode overstable \citep{2015MNRAS.446..240M}. 
To diagnose this behavior in our simulations, we measure how the
Maxwell stress responds in time to pressure perturbations as a function of radius.

Specifically, at each radius, we compute the cross-correlation between the shell-averaged total pressure $P$ and the Maxwell stress $T$. 
For each radius, we calculate the normalized time-domain cross-correlation
\begin{equation}
C_{\rm ps}(r,\tau) =
\frac{\displaystyle\int_{t_1}^{t_2}
        [P(r,t)-\overline{P}]\,[T(r,t+\tau)-\overline{T}]\,{\rm d}t}
     {\sigma_P\,\sigma_T\,(t_2-t_1)},
\end{equation}
where $\overline{P}$ and $\overline{T}$ are time averaged quantities over the time interval $[t_1,t_2]$, and $\sigma_P$ and $\sigma_T$ are the corresponding standard
deviations.
We restrict our attention to positive lags $\tau\ge 0$ and define,
for each radius,
\begin{equation}
\tau_{\rm peak}(r) = \arg\max_{\,0\le\tau\le \tau_{\max}} C_{\rm ps}(r,\tau),
\end{equation}
\begin{equation}
C_{\rm ps,peak}(r) = C_{\rm ps}\bigl(r,\tau_{\rm peak}(r)\bigr),
\end{equation}
where $\tau_{\max}$ chosen as $\sim 10$ local orbital periods. 
In the idealized case of a nearly monochromatic oscillation with frequency $\omega$, the cross-correlation behaves as $C_{\rm ps}\propto \cos(\omega\tau-\phi)$; the lag $\tau_{\rm peak}$.  
Therefore, it provides an effective measure of the stress--pressure phase lag, $\phi\simeq \omega\,\tau_{\rm peak}$. 
While $C_{\rm ps,peak}$ quantifies the strength of the coherent coupling between pressure and stress.

\begin{figure}
    \centering
    \includegraphics[width=0.45\textwidth]{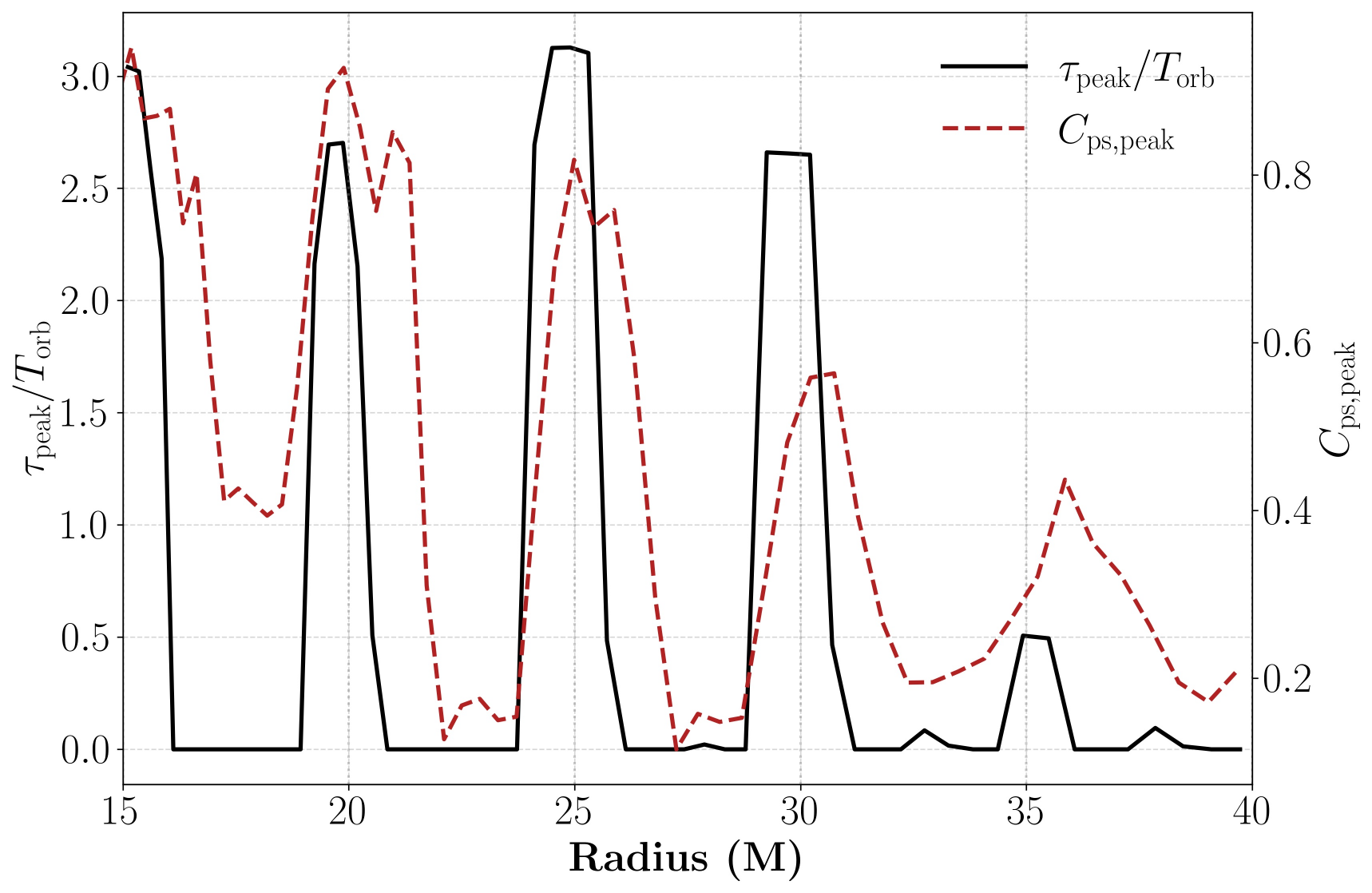}
    \caption{Radial profiles of the pressure-stress lag (black solid) and its correlation (red-dashed) in model~\texttt{Th3D05}.}
    \label{fig:lag}
\end{figure}

Figure~\ref{fig:lag} shows the radial profiles of $\tau_{\rm peak}/T_{\rm orb}$ and $C_{\rm ps,peak}$.
Within each magnetic loop, the cross-correlation peaks at essentially zero lag, and the correlation amplitude remains modest ($C_{\rm ps,peak}\sim 0.1$--$0.3$), indicating that pressure and stress fluctuate almost synchronously and are dominated by local MRI turbulence.
In contrast, at the radii corresponding to the density/magnetization
interfaces—spaced by $\Delta r\simeq 5\,r_g$, matching the loop size---the peak correlation shifts to positive lags of $\tau_{\rm peak}\simeq 2$--$3\,T_{\rm orb}$ and the correlation strength increases to $C_{\rm ps,peak}\gtrsim 0.5$.
These interface radii coincide with the locations of strong QPO power in the PSD maps, suggesting that the interfaces not only act as trapping regions for inertial--acoustic waves but also host a delayed viscous response whereby stress variations lag pressure perturbations over several orbital periods.

This behavior is qualitatively consistent with the viscous-epicyclic
overstability picture, in which a phase-lagged viscous stress extracts energy from differential rotation and feeds inertial-acoustic modes.

\section{Observational implications}

Our simulations have direct implications for the variability properties of stellar-mass black holes during different spectral states, particularly during the decaying phase of outbursts in BHXRBs, where the accretion flow is proposed to evolve from a geometrically thin disk to a thicker, truncated configuration. For a $10\,M_\odot$ black hole, the characteristic frequency scale is $\nu_0 \approx 3.2 \times 10^3$~Hz. Accordingly, at truncation radii $r_{\rm tr} \sim 20$–$40,r_g$, the radial epicyclic frequency corresponds to $\nu_r \sim 20$–$150$~Hz, while the dominant feature at $\sim 3\nu_r$ lies in the range $\sim 60$–$450$~Hz. These values are consistent with the reported HFQPOs in sources such as GRS~1915+105, XTE~J1550–564, GRO~J1655–40, H1743–322, and IGR~J17091–3624 \citep{2006ARA&A..44...49R,2014SSRv..183...43B,2006ApJ...637.1002R,1998astro.ph..6049R,2012ApJ...747L...4A}.   Our simulations further demonstrate that the truncation radius depends strongly on the magnetic field configurations, spanning $r_{\rm tr}  \sim 20$–$60,r_g$ across different loop configurations. For a $10\,M_\odot$ black hole, these truncation radii are consistent with observationally inferred inner disk radii in systems such as GX~339–4 and H1743–322 \citep{2007A&ARv..15....1D,2011MNRAS.415.2323I,2015A&A...573A.120P,2015ApJ...813...84G}. This suggests the possibility of the presence of different configurations in accretion flows around black holes. Larger-scale magnetic loop cases produce more extended puffed-up inner regions and larger truncation radii, while smaller-scale loop cases generate more compact, bulky inner disks, providing natural explanations for different observed sources \citep{2009ApJ...707L..87T,2017MNRAS.472.4220B,2019MNRAS.490.1350B}. Thus, the diversity in the observed truncation radii may reflect differences in magnetic configuration rather than being influenced solely by the accretion rate or radiative efficiency \citep{2015ApJ...809..118B,2024ApJ...966...47L}. This result opens up a new frontier for further investigation in future works. 

A major finding of this study is that viscous-epicyclic overstability operates efficiently only in geometrically thin disk regions. As the accretion disk becomes thicker, enhanced turbulent diffusion and reduced epicyclic restoring forces suppress coherent oscillations, leading to the disappearance of QPOs. This transition is consistent with observations showing that HFQPOs are strongest in intermediate and weaken in hard states where the flow becomes geometrically thicker \citep{2006ARA&A..44...49R,2005A&A...440..207B,2015ApJ...813...84G}. Furthermore, the detection and coherence properties of QPOs offer an important probe of the underlying disk geometry. Specifically, whether the inner flow is geometrically thin or thick, and of its temporal evolution during outbursts. Such transitions from geometrically thin to thick disk are often tracked using the spectral state analysis \citep{2004MNRAS.355.1105F,2011BASI...39..409B}. Together, our study provides a unified interpretation in which magnetic topology, truncation radius, and disk geometry jointly regulate QPO frequencies, coherence, and evolution during the decaying phase of outbursts in BHXRBs.

Although our simulations do not include radiative transfer, the locations of the strongest oscillatory signals allow us to infer where the associated emission would most likely originate. The global $3\,\nu_r$ variability is strongest in the inner, relatively thin portions of the disk and near the truncation region, where dissipation and accretion power are both enhanced. In BHXRBs, these regions are expected to contribute predominantly to the X-ray emission \citep{1973A&A....24..337S,1973blho.conf..343N,2007A&ARv..15....1D}, especially the thermal disk component and Comptonized hard X-ray emission arising from the puffed-up inner flow or hot corona \citep{2006ARA&A..44...49R}. By contrast, the jet-associated radio emission is produced much farther from the black hole and over much longer variability timescales \citep{2004MNRAS.355.1105F,2009ApJ...707L..87T,2017MNRAS.472.4220B}. The high-frequency oscillations reported here are less likely to appear directly in the radio band, except possibly after propagation-induced smoothing and with significant delay. Therefore, our results suggest that the most direct observational signature of these oscillations should appear in X-rays, while lower-frequency or reprocessed counterparts at longer wavelengths may be weaker and less coherent \citep{2014SSRv..183...43B}.

\section{Summary and Discussion} \label{sec:summary}

In this work, we performed a series of GRMHD simulations of geometrically thin accretion disks. By systematically varying the initial magnetic field topology, specifically, the loop size and radial dependence of the multi-loop SANE-type magnetic configuration, we investigated how magnetic geometry regulates the disk’s vertical structure, jet production efficiency, and time-dependent variability. Both 2D and 3D simulations were conducted to ensure the robustness of the observed behavior.

Our main findings can be summarized as follows:

\begin{enumerate}
    \item Magnetic topology controls outflow (jet and wind) power and disk truncation.
    Larger magnetic loops or stronger radial dependence in the initial field configuration yield higher magnetic flux threading the black hole, resulting in enhanced jet and wind power. These configurations also produce broader, magnetically supported inner regions and larger truncation radii, suggesting that magnetic geometry plays a dual role in both energy extraction and vertical disk structure formation. This may imply that the observed spread in inner-disk radii among BHXRBs is set, at least in part, by the global magnetic-field structure of the flow, in addition to the usual dependence on accretion state.

    \item The Low-$\beta$ regions correlate with vertical expansion. 
    The spatial correspondence between the low plasma $\beta$ zones and the enhanced disk scale height indicates that magnetic pressure support and buoyancy drive the vertical puff-up of the inner flow. However, the early onset of vertical expansion and the appearance of inclined stripe-like features in the space-time diagrams cannot be fully explained by the evolution of plasma $\beta$ alone, implying the involvement of additional dynamical processes.

    \item Viscous-epicyclic overstability would be the possible origin of QPOs. 
    Analysis of Maxwell $\alpha$ viscosity shows coherent QPOs whose frequencies match the local epicyclic frequency $\nu_r(r)$. These oscillations arise from the viscous–epicyclic overstability, a self-excited interaction between viscous stress and inertial–acoustic perturbations. The multi-loop magnetic field topology provides natural radial cavities that trap these oscillations, leading to long-lived, stripe-like features in space-time diagrams. This places the simulated variability in a parameter range broadly compatible with the HFQPO behavior reported in several BHXRBs.

    \item 
    As the disk becomes thicker and plasma $\beta$ increases, the epicyclic restoring force weakens, and viscous diffusion damps the overstability. The oscillations gradually lose coherence, leading to the disappearance of organized stripes and the emergence of stochastic turbulence. This transition is consistent with the theoretical expectation that viscous-epicyclic modes operate efficiently only in a geometrically thin disk region. In this framework, changes in QPO strength and coherence can be viewed as signatures of the gradual reorganization of the inner flow during outburst evolution.

    \item 
    Our 3D thin-disk run confirms the 2D results, revealing both global bright bands following the $3\,\nu_r$ curve and local QPOs near the density transition layers at $r \sim 20\,r_g$ and $r \sim 40\,r_g$. These boundary-associated QPOs arise from shear interfaces that act as local resonant cavities, partially reflecting and amplifying inertial–acoustic perturbations. The coexistence of global ($3\,\nu_r$) and local (interface) oscillations suggests a hierarchical coupling between large-scale viscous–epicyclic modes and small-scale shear-induced resonances.

    \item 
    By measuring the cross-correlation between total pressure and Maxwell stress, we identify a distinct dynamical signature. Within magnetic loops, stress and pressure fluctuate synchronously, consistent with standard MRI turbulence. In contrast, the loop interfaces are characterized by sharp radial gradients and localized cavities in the effective viscosity $\alpha_M$. These viscosity features act as resonant traps where the stress develops a finite phase lag relative to pressure. This combination of structural confinement (by viscosity gradients) and delayed response (driving the overstability) provides direct evidence that the observed oscillations are physically distinct from background turbulence.
\end{enumerate}

Overall, our simulations demonstrate that the dynamical evolution and variability of magnetized thin disks are governed by an intricate interplay between magnetic topology, viscous feedback, and epicyclic dynamics. The results provide a unified picture in which global oscillations driven by viscous–epicyclic overstability coexist with local shear-trapped modes, jointly shaping the multi-timescale variability of the accretion flow.
This framework naturally connects the low- and high-frequency QPOs observed in numerical disks—and potentially in BHXRBs—to the interplay between magnetic geometry and viscous feedback. The QPO-producing mechanism we presented here is physically non-trivial and robust at the level of dynamics, but its long-term persistence is not guaranteed, because the magnetic configuration that enables it is unlikely to remain stationary in either simulations or real accretion flow. As the disk becomes thicker and turbulent dissipation increases, the coherence of the oscillation is naturally reduced.

Future work should incorporate radiative cooling and a wider range of black hole spins to assess the observational relevance of these oscillations in realistic accretion environments. Extending this analysis to fully three-dimensional, radiation-GRMHD simulations will be essential to establish whether the viscous-epicyclic overstability can operate in luminous, radiatively efficient systems and to determine its potential role in producing the observed QPOs in BHXRBs.

\begin{acknowledgements}
This research was supported by the National Key R\&D Program of China (grant No. 2023YFE0101200), the National Natural Science Foundation of China (grant No. 12273022 and 12511540053), and the Shanghai Municipality Orientation Program of Basic Research for International Scientists (grant No. 22JC1410600). 
The simulations were analyzed on the TDLI-Astro cluster at Shanghai Jiao Tong University.
\end{acknowledgements}

\begin{appendix}

\section{A Thin disk model}\label{appenA}
\renewcommand{\thefigure}{A\arabic{figure}}

For a setup of the thin disk model, we follow \cite{2021MNRAS.505.3596D,2022MNRAS.517.5032D}, which are based on the standard thin-disk model of \cite{1973blho.conf..343N}. In this setup, the initial density distribution on the poloidal plane in Boyer–Lindquist (BL) coordinates is given by
\begin{equation}
\rho(r,\theta) = \rho_{\rm e}(r)\exp\left(-\frac{\alpha^2 z^2}{H_{\rm e}^2}\right), \qquad z = r\cos\theta.
\end{equation}
To maintain the geometrically thin nature of the initial disk, we choose $\alpha = 2$ and the equilibrium disk height $H_{\rm e}$ following \cite{2022MNRAS.517.5032D}. In the above equation, $\rho_{\rm e}(r)$ provides the density profile on the equatorial plane, given by
\begin{equation}
\rho_{\rm e}(x) = \left(\frac{\Theta_0}{\mathcal{K}}\right)^{1/(\Gamma_{\rm g}-1)} f(x)^{1/(4(\Gamma_{\rm g}-1))},
\end{equation}
where $x = \sqrt{r}$, $\Theta_0$ is a constant that fixes the initial temperature distribution of the thin disk. In this work, we adopt $\Theta_0 = 0.001$ and the entropy constant $\mathcal{K} = 0.1$. The function $f(x)$ is defined as
\begin{align}
f(x) &= \frac{3}{2x^2}\frac{1}{x^2(2a+x^3-3x)}
\Bigg[
x - x_0 - \frac{3}{2}\ln\!\left(\frac{x}{x_0}\right) \nonumber \\
&\quad - \frac{3(l_1-a)^2}{l_1(l_1-l_2)(l_1-l_3)}\ln\!\left(\frac{x-l_1}{x_0-l_1}\right)
- \frac{3(l_2-a)^2}{l_2(l_2-l_1)(l_2-l_3)}\ln\!\left(\frac{x-l_2}{x_0-l_2}\right) \nonumber \\
&\quad - \frac{3(l_3-a)^2}{l_3(l_3-l_1)(l_3-l_2)}\ln\!\left(\frac{x-l_3}{x_0-l_3}\right)
\Bigg],
\end{align}
where $x_0 = \sqrt{r_0}$, and $r_0$ is the radius of the innermost stable circular orbit. The quantities $l_1,\,l_2$, and $l_3$ are the roots of the cubic equation $l^3 - 3l + 2a = 0$, whose explicit forms can be found in \cite{1974ApJ...191..499P}.

To completely specify the initial condition, the azimuthal velocity consistent with the above density profile is given by
\begin{equation}
u^{\phi}(r,\theta) = \left(\frac{\mathcal{A}}{\mathcal{B} + 2C^{1/2}}\right)^{1/2},
\end{equation}
where
\begin{align}
\mathcal{A} &= (\Gamma^{r}_{tt})^2, \nonumber \\
\mathcal{B} &= g_{tt}\left(\Gamma^{r}_{tt}\Gamma^{r}_{\phi\phi} - 2\Gamma^{r}_{t\phi}{}^2\right)
    + 2g_{t\phi}\Gamma^{r}_{tt}\Gamma^{r}_{t\phi} - g_{\phi\phi}(\Gamma^{r}_{tt})^2, \nonumber \\
\mathcal{C} &= \left(\Gamma^{r}_{t\phi}{}^2 - \Gamma^{r}_{tt}\Gamma^{r}_{\phi\phi}\right)
    \left(g_{t\phi}\Gamma^{r}_{tt} - g_{tt}\Gamma^{r}_{t\phi}\right)^2.
\end{align}
Here, $\Gamma^{\alpha}_{\beta\gamma}$ and $g_{\mu\nu}$ denote the non-zero components of the Christoffel symbols and the metric tensor of a Kerr black hole, respectively. 
We employ the \textit{funky modified Kerr–Schild (FMKS)} coordinate system, which provides a smooth mapping from Boyer–Lindquist coordinates while concentrating numerical resolution near the equatorial plane and reducing the Courant time-step constraint around the poles. The transformation from BL to FMKS coordinates is applied before the simulation begins to initialize all GRMHD quantities consistently on the computational grid.
The computational grid covers radii from just inside the event horizon out to $1000\,r_g$, and spans the full polar and azimuthal ranges, $\theta \in [0,\pi]$ and $\phi \in [0,2\pi]$. Here $r_g \equiv GM/c^2$ denotes the gravitational radius. The grid resolution for each model is seen in the table.~\ref{Model comparison}.

\section{Multi-loop magnetic field}\label{appenB}
\renewcommand{\thefigure}{B\arabic{figure}}

The magnetic field configuration we adopt follows \citet{2020MNRAS.495.1549N}. Although this multi-loop setup is used here as an idealized initial condition, such a field geometry may be physically plausible in realistic accretion environments. In particular, previous GRMHD simulations have shown that turbulence driven by the magnetorotational instability and large-scale dynamo action can reorganize initially toroidal magnetic flux into poloidal loops and produce the characteristic ``butterfly''-type field reversals seen in stratified disks \citep{2020MNRAS.494.3656L,2018ApJ...861...24H,2024MNRAS.527.3018Z,2025arXiv251203443Z}. This suggests that magnetic structures with alternating polarity need not be purely artificial, but may arise naturally from the dynamo process of magnetized accretion flows. In addition, stellar winds feeding the accretion flow may carry magnetic fields with different polarities. If so, this could also create a multi-loop structure in the torus \citep{2025ApJ...990...81J,2024A&A...688A..82J,2026A&A...707A..27X}. While this possibility has not yet been explored in detail in studies of wind-fed accretion, it provides a plausible physical motivation for considering multi-loop magnetic topologies. The magnetic field configuration used in our work is given by:
\begin{equation}
    A_\phi \propto \max(\rho - \rho_{\min}, 0)\,
\cos[(N-1)\,\theta]\,
\sin\!\left[\frac{2\pi (r-r_\text{nom})}{\lambda_r}\right]r^w.
\end{equation}
where $\rho_{\min}=0.01$ and $N=3$ for all simulations in this work.
$r_{\mathrm{nom}}$ is defined as the normalized radius corresponding to the maximum density obtained from the thin disk model introduced in Appendix~\ref{appenA}. The amplitude of $A_\phi$ is chosen such that the minimum plasma beta satisfies $\beta_{\min} = 20$, with $\beta = p_{\mathrm{g}} / p_{\mathrm{mag}}$.

\section{Quality factor}\label{appenC}
\renewcommand{\thefigure}{C1}
Here, we investigate whether our simulations can capture the MRI in the disk well.
In practice, the fastest-growing MRI mode is considered well resolved when $Q \gtrsim 6 $ \citep{2004ApJ...605..321S}. 
Figure~~\ref{fig:quality factor} shows MRI quality factors of three directions for model \texttt{Th3D20} at $t=10,000\,M$. It is indicated that the MRI is well resolved (e.g., $Q$ is larger than 10) throughout most of the domain. Therefore, our grid resolution is enough to capture the evolution of simulations properly.

\begin{figure}
    \centering
    \includegraphics[width=0.8\linewidth]{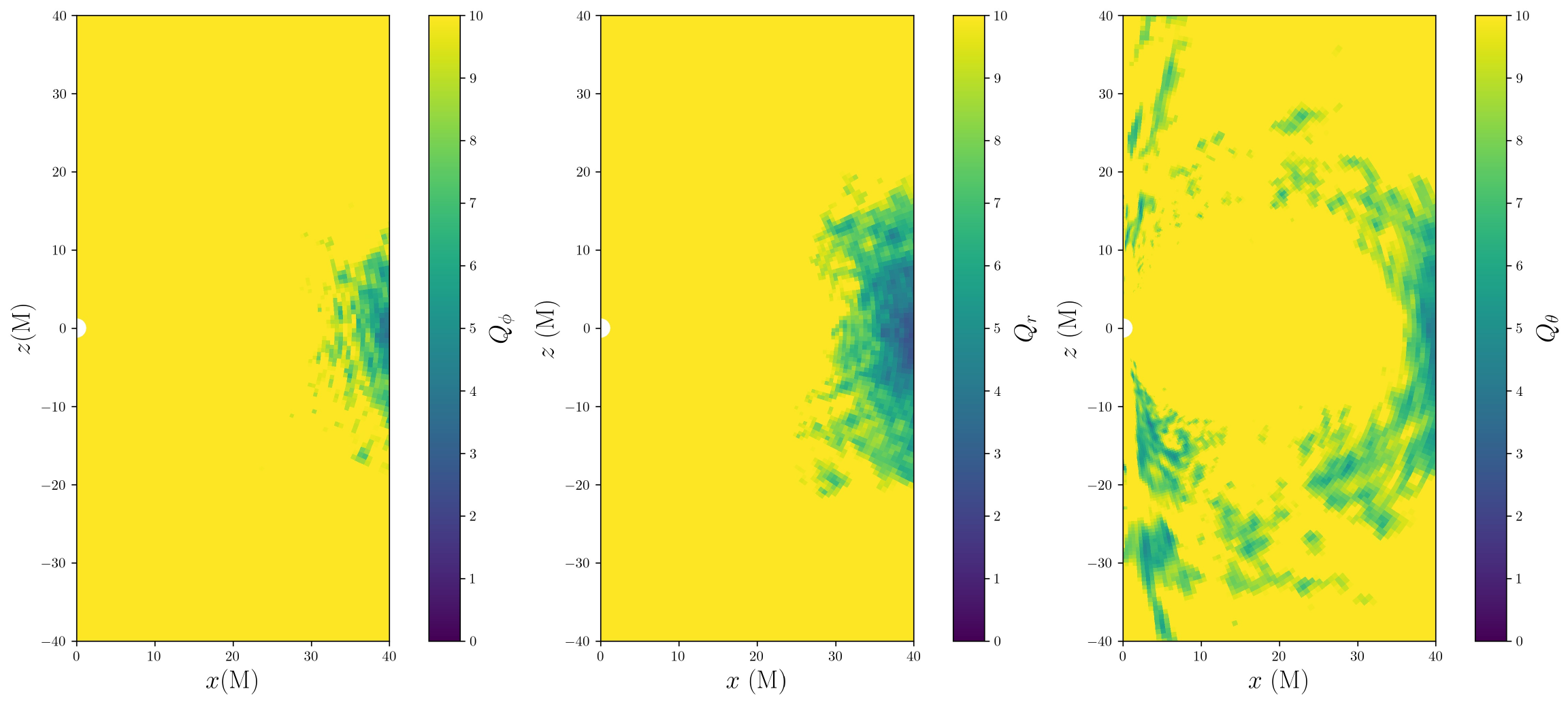}
\caption{MRI quality factor for model $\tt{Th3D20}$ at $t = 10{,}000\,M$. 
}
    \label{fig:quality factor}
\end{figure}

\section{Loop size and the location of the interface}\label{appenD}
\renewcommand{\thefigure}{D1}
To more clearly demonstrate the modulating effect of the loop size on the QPO at the interface, we have carried out additional simulations with a further reduced loop size. Figure~\ref{fig:3D-spacetime-Mdot-05-combined} presents 
the spacetime diagram of the mass accretion rate $\dot{M}$ for the model
\texttt{Th3D05}. Compared to 
Figure~\ref{fig:3D-spacetime-Mdot}, more interfaces associated with the smaller loop are now resolved. Therefore, we can obtain the same conclusion that QPOs arise from the interface of each loop.

\begin{figure}
    \centering
    \includegraphics[width=0.4\linewidth]{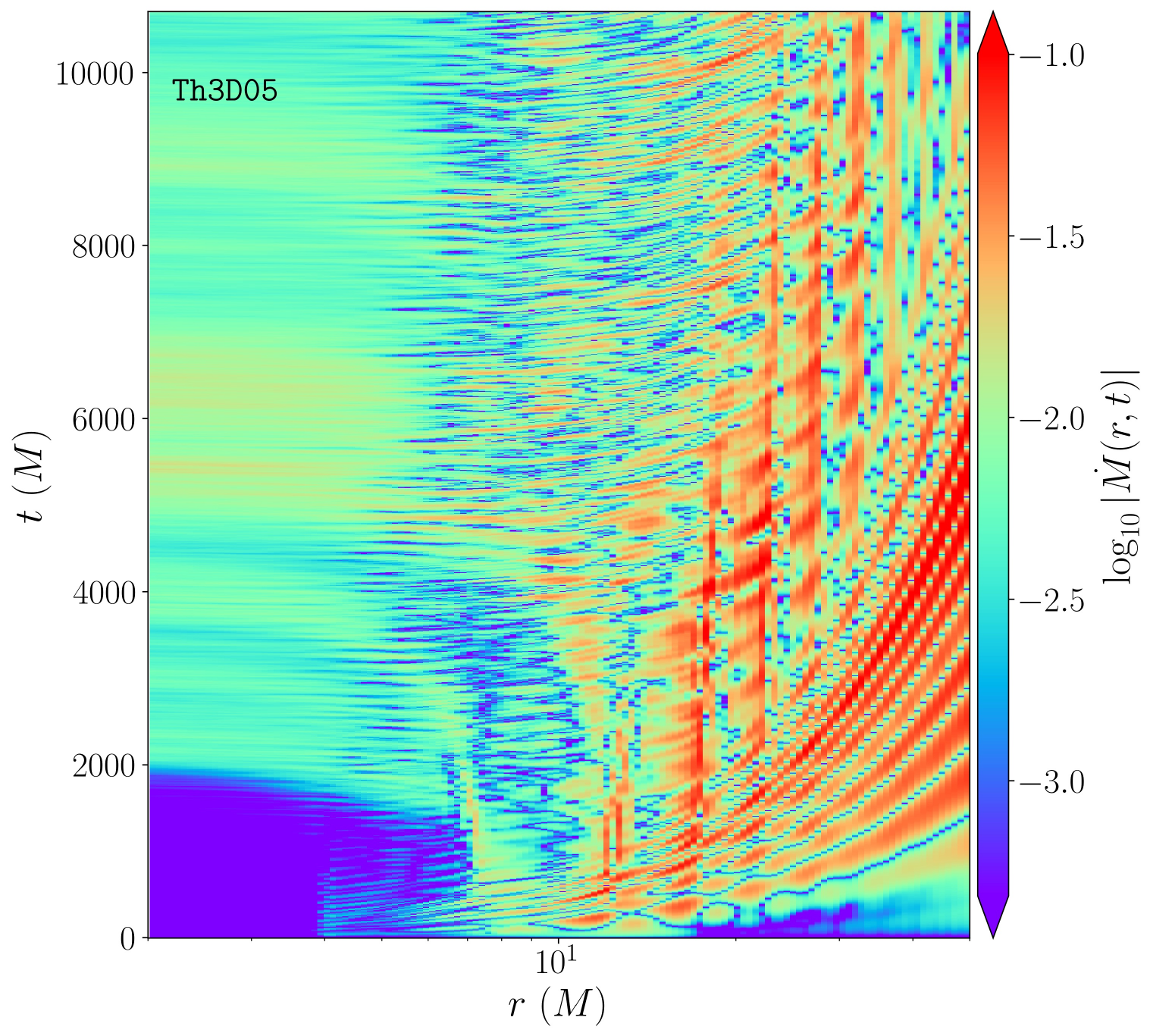}
\caption{Same as Figure~\ref{fig:3D-spacetime-Mdot} but for \texttt{Th3D05}.
}
    \label{fig:3D-spacetime-Mdot-05-combined}
\end{figure}

\end{appendix}

\clearpage
\bibliography{sample701}{}
\bibliographystyle{aasjournalv7}



\end{document}